\pdfoutput=1
\documentclass[10pt,journal,compsoc]{IEEEtran}
\IEEEoverridecommandlockouts

\ifCLASSOPTIONcompsoc
  \usepackage[nocompress]{cite}
\else
  \usepackage{cite}
\fi

\ifCLASSINFOpdf
  \usepackage[pdftex]{graphicx}
\else
\fi
\usepackage[utf8]{inputenc}

\usepackage[para,online,flushleft]{threeparttable}
\usepackage[english]{babel}
\usepackage{amsthm}
\usepackage{amsmath}
\usepackage{varwidth}
\usepackage{blindtext}
\usepackage{multicol}
\usepackage{graphicx}
\usepackage{paralist}
\usepackage[english]{babel}
\usepackage{float}
\usepackage{makecell}
\usepackage{multirow}
\usepackage{amsfonts}
\usepackage{booktabs}
\usepackage{siunitx}
\usepackage{etoolbox}
\usepackage{subcaption}
\usepackage{numprint}
\usepackage[T1]{fontenc}
\usepackage{algorithm}
\usepackage{flushend}
\usepackage[noend]{algpseudocode}
\usepackage{textcomp}
\usepackage{xcolor}
\usepackage{fancyvrb}
\usepackage{paralist}
\usepackage[inline]{enumitem}
\usepackage{booktabs}
\usepackage{color}
\usepackage{colortbl}
\usepackage{flexisym}
\usepackage{array}
\usepackage{xparse}
\usepackage{xspace}
\usepackage[flushleft]{threeparttable}
\usepackage[nobreak]{mdframed}
\usepackage{framed}
\usepackage{pifont}

\newcommand{\cmark}{\ding{51}}%
\newcommand{\xmark}{\ding{55}}%

\ifCLASSOPTIONcompsoc
  \usepackage[nocompress]{cite}
\else
  \usepackage{cite}
\fi

\usepackage{listings}
\usepackage[]{xcolor}

\theoremstyle{definition}
\newtheorem{definition}{Definition}

\newcommand{\toolname}{\textsc{Collector-Sahab}\xspace}
\newcommand{\ddiff}{\textsc{didiffff}\xspace}
\newcommand{\selogger}{\textsc{selogger}\xspace}
\newcommand{\ds}{\textsc{MVN-DRR}\xspace}
\newcommand{\dssize}{\textsc{584}\xspace}
\newcommand{\rqthreeds}{RW3\xspace}

\newcommand{\revision}[1]{\textcolor{black}{#1}}

\usepackage[]{xcolor}
\newcommand{\TODO}[1]{\textcolor{red}{#1}\GenericWarning{}{LaTeX Warning: TODO: #1}}\newcommand\todo\TODO

\newcommand{\GHilight}{\makebox[0pt][l]{\color{green}\rule[-4pt]{1\linewidth}{10pt}}}
\newcommand{\RHilight}{\makebox[0pt][l]{\color{pink}\rule[-4pt]{1\linewidth}{10pt}}}
\newcommand{\YHilight}{\makebox[0pt][l]{\color{yellow}\rule[-4pt]{1\linewidth}{10pt}}}

\algnewcommand{\Inputs}[1]{%
  \Statex \textbf{Inputs:}
  \Statex \hspace*{\algorithmicindent}\parbox[t]{.8\linewidth}{\raggedright #1}
}
\algnewcommand{\Outputs}[1]{%
  \Statex \textbf{Outputs:}
  \Statex \hspace*{\algorithmicindent}\parbox[t]{.8\linewidth}{\raggedright #1}
}
\algnewcommand{\Initialize}[1]{%
  \State \textbf{Initialize:}
  \Statex \hspace*{\algorithmicindent}\parbox[t]{.8\linewidth}{\raggedright #1}
}

\definecolor{javared}{rgb}{0.6,0,0} 
\definecolor{javagreen}{rgb}{0.25,0.5,0.35} 
\definecolor{javapurple}{rgb}{0.5,0,0.35} 
\definecolor{javadocblue}{rgb}{0.25,0.35,0.75} 

\lstdefinestyle{diff}{
    escapechar=\%
}

\newcommand{\collectorsahab}{\textsc{Collector-Sahab}\xspace}

\usepackage{hyperref}
\addto\extrasenglish{%

}

\begin{document}

\title{Augmenting Diffs With Runtime Information}

\author{
Khashayar Etemadi,
Aman Sharma,
Fernanda Madeiral,
Martin Monperrus%

\IEEEcompsocitemizethanks{
\IEEEcompsocthanksitem K. Etemadi, A. Sharma, and M. Monperrus are with the KTH Royal Institute of Technology, Stockholm, Sweden\protect\\
Email: \{khaes, amansha, monperrus\}@kth.se\protect\\
\IEEEcompsocthanksitem F. Madeiral is with the Vrije Universiteit Amsterdam, Amsterdam, Netherlands\protect\\
Email: fer.madeiral@gmail.com
}
}

\IEEEtitleabstractindextext{
\begin{abstract}
Source code diffs are used on a daily basis as part of code review, inspection, and auditing. To facilitate understanding, they are typically accompanied by explanations that describe the essence of what is changed in the program. As manually crafting high-quality explanations is a cumbersome task, researchers have proposed automatic techniques to generate code diff explanations. Existing explanation generation methods solely focus on static analysis, i.e., they do not take advantage of runtime information to explain code changes. In this paper, we propose \toolname, a novel tool that augments code diffs with runtime difference information. \toolname compares the program states of the original (old) and patched (new) versions of a program to find unique variable values. Then, \toolname adds this novel runtime information to the source code diff as shown, for instance, in code reviewing systems. \revision{As an evaluation, we run \toolname on \dssize code diffs for Defects4J bugs and find it successfully augments the code diff for 95\% (555/\dssize) of them.} 
\revision{We also perform a user study and ask eight participants to score the augmented code diffs generated by \toolname. Per this user study, we conclude that developers find the idea of adding runtime data to code diffs promising and useful. }
Overall, our experiments show the effectiveness and usefulness of \toolname in augmenting code diffs with runtime difference information. \\\textbf{Publicly-available repository:} \revision{\url{https://github.com/ASSERT-KTH/collector-sahab}.}
\end{abstract}

}

\maketitle

\IEEEdisplaynontitleabstractindextext

\IEEEpeerreviewmaketitle
\section{Introduction}

A software program evolves based on a series of changes to its source code. 
Developers are the gatekeepers of this evolution.
Typically, they read, analyze, and ensure the quality of the code difference (hereafter, code diff) between two versions of a program, a best practice known as code review \cite{sadowski2018modern}. 
\autoref{lst:matched_line} shows an example code diff that fixes a bug in Java, with the typical highlighting provided by IDEs and code review systems.

To facilitate the code reviewing process, various forms of explanations may be added to code diffs, such as commit messages \cite{jiang2017automatically}, code comments, and pull request descriptions \cite{liu2019automatic}. As manually crafting a high-quality explanation is time-consuming and may be neglected by developers \cite{liu2018neural,dyer2013boa,liu2021identifying}, researchers have proposed techniques to automatically generate code diff explanations.  Currently, explanation generation mostly focuses on static information, that is extracted from the code change without running the program \cite{jiang2017automatically,liu2018neural,liu2020atom}.
However, runtime information is a great source of data for explaining code diffs \cite{galenson2014codehint}. For example, to properly understand the change in \autoref{lst:matched_line}, developers may need to know the impact of the condition change on the runtime value of variable \texttt{numericalVariance}. 

In this paper, we present a study on using runtime information for explaining code diffs.
This is a challenging task because there is an overwhelming number of events and values happening at runtime.
This endeavor requires three components:
first, a proper algorithm to monitor and select data during the execution of both the original and patched versions of a program;
second, an efficient algorithm to extract runtime differences between the collected traces;
third, the extracted runtime differences should be integrated into code diffs in a useful manner for code reviewers.

\begin{lstlisting}[float,style=diff, caption={A typical diff for a bug-fixing code diff, with state-of-the-art highlighting as found e.g. on Github.}, label=lst:matched_line,language=Java,belowskip=-0.4\baselineskip]
public double getNumericalVariance() {
%\RHilight%-   if (!numericalVarianceIsCalculated) {
%\GHilight%+   if (!(sampleSize <= numberOfSuccesses)) {
      numericalVariance =  calculateNumericalVariance();
      numericalVarianceIsCalculated = true;
    }
    return numericalVariance;
}
\end{lstlisting}


In this paper, we propose \toolname, a novel system to augment code diffs with runtime information.
\revision{Given a code diff and a test case written in Java, \toolname runs the test on both versions and uses Java bytecode instrumentation to collect \textit{program states} during the execution, where program states consist of values assigned to specific, relevant variables.} After collecting program states, \toolname employs an original algorithm to find the variable values that are unique, in the sense that they only occur in one execution of the program under test. The first relevant unique variable value is added to the code diff to obtain an \emph{augmented diff}, that is a diff with both static and dynamic differences. \toolname is the first tool that augments code diffs with relevant runtime difference information.

To assess the effectiveness of \toolname, we run it on \dssize code diffs from the bug benchmark Defects4J \cite{just2014defects4j}.
\revision{Our results show that \toolname detects a unique program state for 95\% (555/\dssize) of the code diffs in the benchmark. This outperforms the most related work, \ddiff \cite{kanda2022didiffff}. \ddiff is a web application to view the runtime value differences of a Java program. To assess \toolname's applicability, we perform a field study on 50 code diffs from a diverse set of 10 real-world open-source projects. \toolname successfully analyzes 96\% (48/50) of these code diffs. We conduct a manual analysis on 30 of those real-world code diffs for which \toolname augments the diff with runtime differences.} This manual analysis confirms the usefulness of \toolname in terms of correctness, understandability, and causal relation with the intended runtime behavioral change.
\revision{Finally, we perform a user study  and ask eight participants to rate them in terms of their usefulness, clarity, and novelty. The results confirm that developers find augmenting code diffs with runtime data a useful and promising practice.}

To sum up, \toolname is a novel tool that goes beyond existing explanation tools that only focus on static data \cite{liu2020atom}, by integrating runtime differences into code diffs. \toolname is the first approach that can provide developers with a concise set of runtime differences, usable for code review.

To summarize, we make the following contributions:
\begin{itemize}
    \item We introduce \toolname, a new system for augmenting code diffs with runtime information based on advanced execution monitoring, effective runtime data differencing, and well-integrated rendering for developers. \revision{\toolname is publicly available at \url{https://github.com/ASSERT-KTH/collector-sahab}.}
    \item We report original results of an evaluation of \toolname's effectiveness in capturing runtime differences for \dssize code diffs in Java. The results show that \toolname reports a runtime difference for 95\% (555/\dssize) of the code diffs. \toolname outperforms \ddiff, the state-of-the-art tool in this domain, which only reports differences for 64.2\% (377/\dssize) of the cases.
    \revision{\item We perform a manual analysis on runtime differences detected by \toolname for 30 code diffs from real-world, complex projects.} This analysis confirms that \toolname augmented diffs are correct, understandable, and successfully exclude random runtime differences.
    \revision{\item We report the results of an original user study on code diff augmentation with runtime data. The results of this study and our interview with the participants show that developers find \toolname helpful.}
\end{itemize}

The rest of the paper is organized as follows. In \autoref{sec:terminology}, we review the foundational concepts used in this paper. In \autoref{sec:technical-section}, we explain how \toolname works. \autoref{sec:experimental-design} and \autoref{sec:experimental-results} present the protocol that we use for our experiments and their results. \autoref{sec:threats} discusses the threats to the validity of our results. In \autoref{sec:related}, we review the related work. Finally, in \autoref{sec:conclusion} we conclude the paper.

\section{Core Concepts}
\label{sec:terminology}

In this section, we review three foundational concepts of our paper: \emph{program state}, \emph{state depth}, and \emph{breakpoint}.

The first core concept that is used in this work is program state. This concept is approached in different forms in previous work. Zeller defines program state as the set of program variables and their values \cite{zeller2002isolating}, which is used by other researchers as well \cite{pham2017learning}. There are also more broad definitions for program states. For example, Xu et al. \cite{xu2007efficient} also consider the call stack and program counter to be a part of the program state. Meinicke et al. \cite{meinicke2019variational} take heap objects as a part of the program state as well.

In the literature, there also exist program state definitions that are more narrow and tailored to the task under consideration. For example, the set of variables considered in the program state can be limited to variables used at a particular location \cite{huang2007automated,gupta2020deep}, can be fixed \cite{schaf2013explaining}, can be in-scope variables at a particular execution point \cite{xie2004checking}, can be variables assigned at a certain point in program \cite{abou2018substate}, or finally variables that are read or written before a certain location \cite{mehne2018accelerating}. Considering those various definitions from the literature, we attempt to generically define ``program state'' as follows.

\begin{definition}[Program State -- Generic]
A program state \emph{s} is the set of all variables \emph{$V_l$} and their values accessed at a certain line \emph{l} of a program \emph{p} that is being executed.
 \end{definition}

There are two points to be noted regarding our definition of program state. First, since a line of a program may be executed multiple times during one single execution, it is possible to collect multiple program states for a given program line. Secondly, the read or written variables can refer to either primitive or non-primitive datatypes. This means the program state may include objects. The boundary of what constitutes an object at runtime is ambiguous. To fully capture the content of an object, we have to go inside the object and read the content of its fields, recursively. This leads us to the definition of state depth, as follows. 

\begin{lstlisting}[float=tb, style=diff, caption={A program state is the runtime information associated to some datatype. In this example, the ``Student'' datatype is shown together with an object of this class ``student1'' with its content. The student object contains more or less information, depending on the observation depth, in color.}, captionpos=b, label=lst:object_depth_ex,belowskip=-0.4\baselineskip]
%{\color{blue} \textbf{Datatypes in Java-like syntax:}}%
class Student {
    String name;
    Supervisor supervisor;
}
class Supervisor {
    String name;
    Education education;
}
class Education {
    String university;
    String city;
}
%\hrule%
%{\color{blue} \textbf{An object of the student datatype, serialized in JSON-like syntax:}}%
{
%\GHilight%  "student1": {                                  %{$depth \geq 1$}%
%\GHilight%    "name": "Alice",                             %{$depth \geq 1$}%
%\RHilight%    "supervisor": {                              %{$depth \geq 2$}%
%\RHilight%        "name": "Bob",                           %{$depth \geq 2$}%
%\YHilight%        "education": {                           %{$depth \geq 3$}%
%\YHilight%            "university": "KTH Institute",       %{$depth \geq 3$}%
%\YHilight%            "city": "Stockholm"                  %{$depth \geq 3$}%
        }
    }
  }
}
\end{lstlisting}

\begin{definition}[State Depth]
A program state \emph{s} with depth \emph{d} collected at line \emph{l} consists of all values for primitive variables inside objects accessed at \emph{l} that can be reached with at most \emph{d} steps.\footnote{In our prototype implementation for Java, we consider the nine following datatypes as primitive: int, byte, short, long, float, double, boolean, char, and string.}
\end{definition}

For example, consider \autoref{lst:object_depth_ex}, which shows a class ``Student'' and an object of that class that the variable ``student1'' refers to. The Student class contains a student's name, her supervisor's name, and her supervisor's education information. The object ``student1'' is serialized in JSON-like format at the bottom of the listing. If the state depth is set to 3, we collect all the information inside the object presented in \autoref{lst:object_depth_ex}. If depth is set to 2, we do not collect the data inside ``student1.supervisor.education''. If depth is set to 1, we do not collect the data inside ``student1.supervisor''. Finally, if depth is set to 0, we do not collect any data inside ``student1''. In fact, when depth is set to 0, the program state only includes the value of primitive variables and excludes non-primitive objects such as ``student1''.

To collect the program states during execution, we have to specify the lines at which the states should be collected. 
This corresponds to the well-known debugging concept of ``breakpoint''.
In debugging, breakpoints are used to suspend the program at a certain point of execution and explore the program state \cite{zhang2013automated}. 

\begin{definition}[Breakpoint]
A breakpoint is a certain line \emph{l} of a program \emph{p} to be watched during an execution of \emph{p}.
\end{definition}

In this paper, we use breakpoints to collect program states, without human intervention.
\revision{As opposed to manual debugging, no developer is involved.} When a breakpoint is reached, the program execution is resumed right after the automated program state collection.

\section{Design and Implementation}
\label{sec:technical-section}

\subsection{Overview}
\label{sec:design_overview}

\begin{figure*}
\begin{center}
\includegraphics[width=\textwidth]{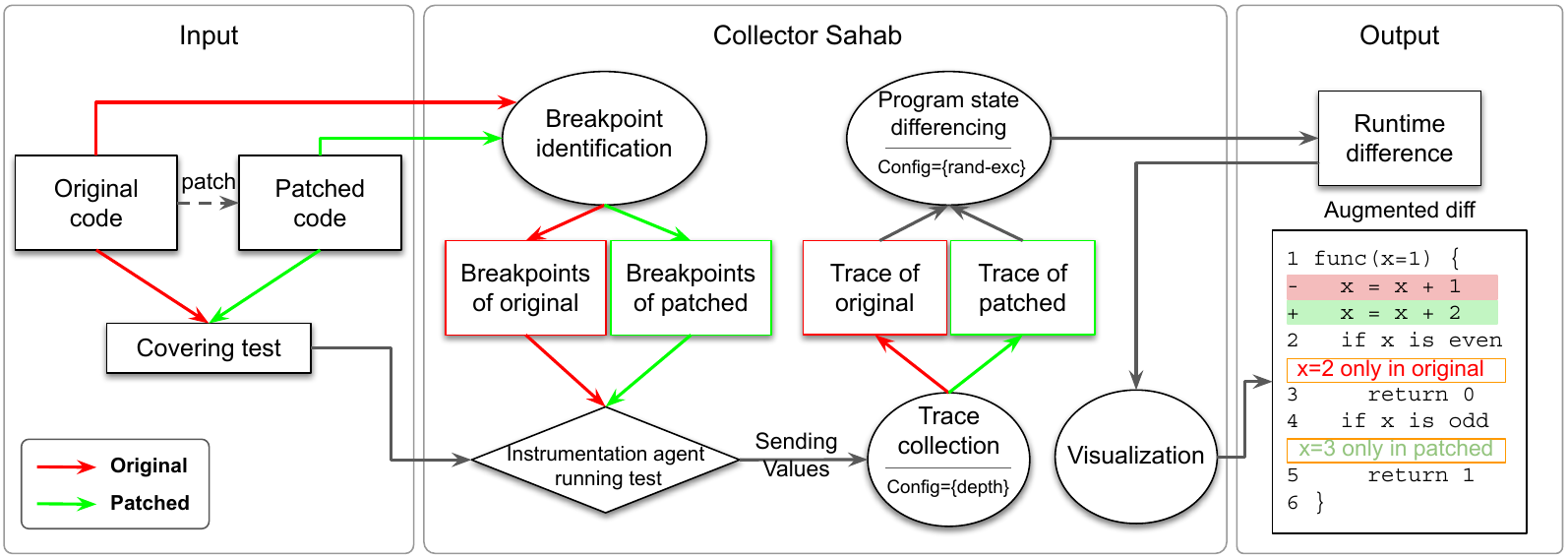}
\caption{\revision{Overview of \toolname.}}
\label{fig:collector-sahab}
\end{center}
\end{figure*}

We design and implement \toolname to generate a concise and useful report of runtime differences caused by a code change. This report puts the extracted execution data into a user interface (UI) to present code diffs. We call this report the \textit{augmented diff} as it augments the code diff with dynamic information. \autoref{fig:collector-sahab} represents an overview of how \toolname works.

\toolname gets an original and patched version of a program as well as a test case to execute them \autoref{sec:appraoch_input}. In the first step, \toolname executes the test on the original and patched versions of the program and compares the program states that occur during these two executions. To collect the program states, \toolname first identifies breakpoints where the data should be collected (see \autoref{sec:breakpoint_identification}) and collects the execution traces at those lines (see \autoref{sec:trace_collection}).

Then, \toolname selects and reports a set of relevant values that occur in the program states of the original version but are absent in the patched version, and vice versa (see \autoref{sec:program_state_diffing}). This set of values represents the relevant runtime differences between two versions of a program. We focus on the relevant values, to make the data presented to code reviewers more precise \cite{santelices2013quantitative}.

Identifying values that occur only in program states of one of the versions can require excessive resources, as the number and size of collected program states can be very large. To address this issue, \toolname adopts an effective algorithm to extract the fine-grained data in all program states and compare them against each other in massive numbers, see \autoref{sec:program_state_diffing}.

Finally, we integrate the extracted relevant execution differences into the code diffs typically used in code review (see \autoref{sec:output}). The result is an augmented diff, it enables developers to see runtime differences, together with the code change.


\subsection{Input}
\label{sec:appraoch_input}

Our tool takes in 3 inputs: the original version of the project, the patched version of the project, and a list of \textbf{covering tests}. A covering test is a test that runs at least part of the code that is changed.
The challenge is to collect enough data and filter the interesting bits when running a covering test on the original and patched versions.

\subsection{Breakpoint Identification}
\label{sec:breakpoint_identification}

\revision{Given the original and patched versions, \toolname first statically analyzes the code to identify the lines of the program that should be monitored as breakpoints. Next, a dynamic analysis is performed on these lines to collect the execution data that is used for runtime difference detection.} For this, \toolname finds a mapping between the lines of the original and patched versions. We call them the \emph{matched lines} as defined below.

\begin{definition}[Diff and Unchanged Lines]
\revision{Given an original and a patched version of a program source file, \emph{diff} is the set of lines that Myer's \cite{Myers1986-dw} algorithm detects as changed lines between the original and patched versions, as implemented in git diff.} The diff contains a set of code lines deleted from the original and a set of code lines inserted into the patched. The remaining lines are the \emph{unchanged lines}.
\end{definition}

\begin{definition}[Intra-function Matched Lines]\label{def:matched_lines}
In this paper, matched lines are all tuples $<l_o,l_p>$ where $l_o$ and $l_p$ meet two requirements. First, $l_o$ is a line from the original version that is unchanged and mapped to $l_p$ from the patched version. Second, $l_o$ should be from a method that contains at least one line of the diff.
\end{definition}

For example, \autoref{lst:matched_line} shows an example of a patch with diff at lines 2 and 3. Lines 4, 5, 6, 7, and 8 are matched lines, for this example. Each of these lines is unchanged in the code diff and appears both in the original ($l_o$) and patched ($l_p$) versions. Also, these lines all appear in the \texttt{getNumericalVariance()} method which contains lines 2 and 3 as the diff. Therefore, lines 4, 5, 6, 7, and 8 meet both requirements and form the set of \emph{matched lines}.

\toolname registers breakpoints at all matched lines in the changed method. We focus on the matched lines so that we can compare the data between the two versions.

\revision{In theory, the runtime difference can be observed only after the change. However, \toolname also monitors the lines before the change for two reasons.
First, it enables us to better exclude random changes, see \autoref{sec:program_state_diffing}.
Second, we note that the changed method can be executed multiple times; hence, a program state difference at a changed line can affect what happens above a code change in the method at later executions.}

\subsection{Trace Collection}
\label{sec:trace_collection}

\begin{lstlisting}[float,caption={Example runtime data obatined with \toolname's trace collection component.}, label=lst:example_output, belowskip=-0.4\baselineskip]
{
  "breakpoint": [
    {
      "file": "foo/BasicMath.java",
      "lineNumber": 5,
      "stackFrameContexts": [
        {
          "positionFromTopInStackTrace": 1,
          "location": "foo.BasicMath:5",
          "stackTrace": [
            "add:5, foo.BasicMath",
            "test_add:11, foo.BasicMathTest"
          ],
          "runtimeValueCollection": [
            {
              "kind": "LOCAL_VARIABLE",
              "name": "x",
              "type": "int",
              "value": 23,
              "fields": null,
              "arrayElements": null
            },
            {
              "kind": "LOCAL_VARIABLE",
              "name": "y",
              "type": "int",
              "value": 2,
              "fields": null,
              "arrayElements": null
            }
          ]
        }
      ]
    }
  ]
}
\end{lstlisting}

After registering the breakpoints, \toolname runs the covering test on both original and patched versions and collects the execution trace as defined below.

\begin{definition}[Program State -- in \toolname)]
The program state \emph{s} that \toolname collects once a breakpoint \emph{b} at line \emph{l} is executed consists of the set of all variables $V_l$ visible at \emph{l} and their values, as well as the method calls leading to the execution of \emph{b}. In practice, the program state object that \toolname collects contains the following information:
\begin{itemize}
    \item \textit{file} is source file where \emph{b} is located.
    \item \textit{lineNumber} is the line number where \emph{b} is located.
    \item \textit{stack frame context} includes : the \textit{stack trace} and the \textit{runtime value collection} as follows.
    \item \textit{stack trace} is the list of program lines that called a series of methods before reaching \emph{b} during the execution.
    \item \textit{runtime value collection} is the set of all variables and their values that are visible when the execution reaches \emph{b}. This includes all local variables and class fields visible at \emph{b}. If the variable is an array, all of its elements are also collected, and if it is a non-primitive variable, all the fields inside the object that it refers to are also stored. We note that \toolname considers strings as primitive variables.
\end{itemize}
\end{definition}

\revision{Note that \toolname considers all visible variables at a line to compute the program state. This means even the variables that refer to objects of classes in third-party libraries are monitored. Therefore, \toolname also detects runtime differences involving library objects.}

\begin{definition}[Execution Trace]
\revision{In this paper, an execution trace is defined as the sequence of all program states that \toolname collects during the execution of a test covering a change. Note that since every inserted breakpoint can be executed multiple times, the trace may include one or more program states per breakpoint.}
\end{definition}

\revision{The collection of execution traces is done as follows. Before the initiation of test execution, we instrument the Java classes to collect the local variables, field, and return values at breakpoints. During the execution of the tests, the instrumented code is invoked and we save the runtime value as Java objects in memory. To avoid exceeding memory, the values of variables are collected only up to a state depth that is given as part of \toolname's configuration (see \autoref{sec:terminology}). Eventually, when the test execution finishes, we output a serialized version of program states to disk. The program states are stored in the output file in the same order as they occur during the execution of the program.}

\autoref{lst:example_output} shows the execution trace collected during the execution of test \texttt{BasicMathTest::test_add}. In this example, one program state is collected. This program state is for line 5 of the file \texttt{foo/BasicMath.java}. There are two variables whose values are stored in this trace: \texttt{x=23} and \texttt{y=2}.

\subsection{Program State Differencing}
\label{sec:program_state_diffing}

\begin{algorithm}[!tb]
\caption{Algorithm for program state differencing.}\label{alg:tool_algorithm}
\small
\begin{algorithmic}[1]
\Inputs{\textit{ot:} The trace collected for the original version \\ \textit{pt:} The trace collected for the patched version \\ \textit{osrc:} The source of the original version \\ \textit{psrc:} The source of the patched version}
\Outputs{\textit{ousv:} List of unique relevant state values in the original trace \\ \textit{pusv:} List of unique relevant state values in the patched trace}

\State $oes \gets program\_states(ot)$ \label{algo:getOriginalBreakpoints}
\State $pes \gets program\_states(pt)$ \label{algo:getPatchedBreakpoints}
\State $osv \gets \textproc{get\_state\_values}(osrc,oes)$ \label{algo:getOriginalStates}
\State $psv \gets \textproc{get\_state\_values}(psrc,pes)$ \label{algo:getPatchedStates}
\State $ousv \gets \textproc{get\_unique\_states}(osv,psv)$ \label{algo:getOriginalUnique}
\State $pusv \gets \textproc{get\_unique\_states}(psv,osv)$ \label{algo:getPatchedUnique}

\State \Return $(ousv,pusv)$ \label{algo:return}
\\

\Function{get\_state\_values}{$src,ess$} \label{algo:getStateValues}
    \State $SV \gets []$
    \For {each execution\_state $es$ in $ess$} \label{algo:sv_for1}
        \State $line \gets es.line$ \label{algo:getLine}
        \State $rv \gets relevant\_vars(src,line)$ \label{algo:getRelevantVars}
        \For {each variable\_data $vd$ in $es$}
            \If{$rv.contains(vd.name)$} \label{algo:relevantCheck}
                \State \begin{varwidth}[t]{\linewidth} 
                $cur\_SV \gets \textproc{extract\_svs}(line,$ \par
                \hskip\algorithmicindent 
                $"",vd)$
                \end{varwidth} \label{algo:call_extractSVS}
                \State $SV.insert(cur\_SV)$
            \EndIf
        \EndFor
    \EndFor
    \State \Return $SV$
\EndFunction

\Function{extract\_svs}{$line,prefix,vd$} \label{algo:extractSVS}
    \State $SV \gets []$
    \If{$is\_primitve(vd)$} \label{algo:addPrimitiveSVStart}
        \State \begin{varwidth}[t]{\linewidth}
        $sv\_id=$ \par 
        \hskip\algorithmicindent $line+prefix+vd.name+vd.value$
        \end{varwidth}
        \State $SV.insert(sv\_id)$
        \State \Return $SV$ \label{algo:addPrimitiveSVEnd}
    \EndIf
    \For {each field\_data $ivd$ in $vd$} \label{algo:extractSVSFor}
        \State \begin{varwidth}[t]{\linewidth}
        $SV.insert(\textproc{extract\_svs}(line,rv,$ \par 
        \hskip\algorithmicindent $prefix+vd.name,ivd))$
        \end{varwidth} \label{algo:recursiveCallExtractSVS}
    \EndFor
    \State \Return $SV$ \label{algo:extarctSVSReturn}
\EndFunction

\Function{get\_unique\_states}{$left\_sv,right\_sv$}
    \State $SV \gets []$
    \For {each state\_value $sv$ in $left\_sv$}
        \If{$!right\_sv.contains(sv)$}
            \State $SV.insert(sv)$
        \EndIf
    \EndFor
    \State \Return $SV$
\EndFunction

\end{algorithmic}
\end{algorithm}

We now explain how \toolname computes the runtime difference between two versions.
We devise a novel algorithm whose overarching goal is to produce a concise runtime difference that makes sense to developers. 
For this, we first define our key concept of \emph{state value}, before its qualified version as relevant state value, and unique relevant state value.

\begin{definition}[State value]
For a given program state \emph{s} corresponding to a breakpoint \emph{b} located at line \emph{l}, a state value is a tuple  $<l,path,value>$. In this tuple, $l$ is the line of \emph{b}, $path$ is the path to a primitive variable inside \emph{runtime value collection} of \emph{s}, and $value$ is the value of that primitive variable.
\end{definition}

The \emph{path} in a state value specifies how a primitive variable can be accessed at the given line. Path appears in form of a list of variable names concatenated with dots such as $v_1.v_2...v_n$. $v_1$ is an object accessible at \emph{l}, $v_n$ is a primitive variable, and $v_i$ is a field in the object referred by $v_{i-1}$. Note that we only consider paths primitive variables in state values. The value for a non-primitive variable is an object reference which is not understandable for developers, hence we do not take them as a part of the state value.

State value is the most fine-grained data that can be extracted from an execution trace. We compute all state values inside the collected execution trace to compute the runtime difference.

In \autoref{lst:object_depth_ex}, if ``student1'' is the variable whose value is collected in program state \emph{s} at a breakpoint at line \emph{l}, the following are the state values that we extract for \emph{s}:

\vspace{5pt}
\small

\noindent <l,student1.name,``Alice''>

\noindent <l,student1.supervisor.name,``Bob''>

\noindent <l,student1.supervisor.education.university,``KTH Institute''>

\noindent <l,student1.supervisor.education.city,``Stockholm''>

\normalsize
\vspace{5pt}

\begin{definition}[Relevant State value]
A state value $<l,var_1...,val>$ is considered \emph{relevant}, if and only if $var_1$ is accessed on line \textit{l}.

\end{definition}

\begin{definition}[Unique Relevant State value]
A state value is said to be unique if it only appears on the original or patched version. 
$<l,var.**,val>$ is a unique relevant state value if it is a relevant state value in the original version, and $<l',var.**,val>$ is not a relevant state value in the patched version where $l'$ is the line mapper to $l$ in the matched lines. Conversely, $<l',var.**,val>$ is a unique relevant state value if it is a relevant state value in the execution trace of the patched version and $<l,var.**,val>$ is not among the relevant state values of the execution trace of the patched version.
\end{definition}

Algorithm \autoref{alg:tool_algorithm} describes the algorithm we employ to compute the execution differences. The inputs to the algorithm are the collected traces for the original (\textit{ot}) and patched (\textit{pt}) versions as well as the original (\textit{osrc}) and patched (\textit{psrc}) versions of the source code. The algorithm outputs the unique relevant state values for the original (\textit{outsv}) and patched (\textit{pusv}) versions.

The algorithm first extracts the program states saved in collected traces for each version (lines \autoref{algo:getOriginalBreakpoints} and \autoref{algo:getPatchedBreakpoints}). Note that each collected trace contains the list of all program states that the trace collector has encountered while executing the matched lines. These program states are saved in the same order as they occur during the execution.

Next, all state values are computed for each version (lines \autoref{algo:getOriginalStates} and \autoref{algo:getPatchedStates}). The \textproc{get\_state\_values} method at line \autoref{algo:getStateValues} traverses program states one by one (line \autoref{algo:sv_for1}). For each state, it gets the program line where that state has happened (line \autoref{algo:getLine}) and the relevant variables at that program line (line \autoref{algo:getRelevantVars}). Note for each line, the relevant variables are the ones that are accessed to be read on that line. Then, for each variable data \textit{vd} that corresponds to a relevant variable (see line \autoref{algo:relevantCheck}), we get all state values linked to it by calling the \textproc{extract\_svs} method (line \autoref{algo:call_extractSVS}).

The \textproc{extract\_svs} method (line \autoref{algo:extractSVS}) discovers all state values that can be extracted from a given variable, recursively. For a given primitive variable, the algorithm adds its variable name and value to the state values set (lines \autoref{algo:addPrimitiveSVStart}-\autoref{algo:addPrimitiveSVEnd}) and returns it. For non-primitive variables, the algorithm looks into the fields of the object that the variable is referring to one by one (line \autoref{algo:extractSVSFor}). Per each field in the referred object, \textproc{extract\_svs} is called recursively (line \autoref{algo:recursiveCallExtractSVS}). When the state values are extracted for all the fields of the referred object, the resulting collection of state values is returned (line \autoref{algo:extarctSVSReturn}).

After running lines \autoref{algo:getOriginalStates} and \autoref{algo:getPatchedStates}, the state values that occur in the original and patched versions are saved in \textit{osv} and \textit{psv}, respectively. To compute the unique relevant state values for each version, we iterate over the extracted state values and identify the ones that are not included in the state values of the opposite version. The \textproc{get\_unique\_states} method that is called on lines \autoref{algo:getOriginalUnique} and \autoref{algo:getPatchedUnique} does exactly this. The result of calling this method is then saved in \textit{ousv} and \textit{pusv}, which form the returned result of the program state differencing algorithm.

Note that \revision{\autoref{alg:tool_algorithm}} explains how unique state values are extracted. \toolname also computes unique program states in three steps. First, it maps each program state of each execution trace to the hashed version of that program state. For this, the Java string hashing algorithm is used. Second, the hashes are compared to each other to detect unique program state hashes. Finally, \toolname looks back to the mappings to find the original program states mapped to the detected unique hashes. The result is the list of unique program states in each of the original/patched execution traces.

\revision{As \autoref{fig:collector-sahab} shows, \toolname also has a \texttt{random-exclusion} configuration for the program state differencing state, noted as \texttt{Config={rand-exc}} in the figure. When \texttt{random-exclusion} is set to \texttt{true}, \toolname excludes state values that show random behavior, i.e. not having identical values accross runs. Using this setting enables the users to focus on runtime differences that are causally related to the code changes. For this purpose, \toolname takes two steps. First, it does not consider values of variables that refer to file paths. This step is taken because our experiments on real-world commits show that, most of the time, file paths change on different system environments. This means a runtime difference in file paths is usually because of a change in the environment, not a change in the logic of the program. In the second step, \toolname runs each version of the program three times and only keeps state values that appear in all three runs. This step excludes values that reflect flaky behavior of a program. Excluding random state values generates a code diff augmentation that shows the crucial behavioral changes caused by the code changes.}

\subsection{Output \& Visualization}
\label{sec:output}

\begin{figure}
\begin{center}
\includegraphics[width=\columnwidth]{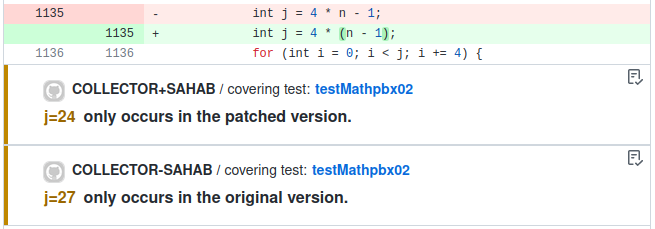}
\caption{The augmented diff generated by \toolname for a code diff for Math-80, showing important runtime information.}
\label{fig:UI_Example}
\end{center}
\end{figure}

The output of the tool is a \textbf{runtime difference} in two forms: a textual file and a visualization. The textual file contains all unique program states and unique state values. It is intended to be used by other tools, i.e. it is not user-facing. For example, such an output can be used by program repair tools to suggest non-overfitting patches based on runtime data.
The other output is aimed at developers, it is a visualization of the runtime difference (see \autoref{fig:UI_Example}). This visualization, which we call the \emph{augmented diff}, is only generated when a unique state value is identified, and detecting a unique program state is not sufficient to produce an augmented diff.

\emph{Textual Output.}
The main textual output of \toolname is a file containing all unique program states and unique relevant state values for both the original and patched versions. Also, all the traces collected at the trace collection step (\autoref{sec:trace_collection}) are stored in a separate file. These textual outputs can be consumed by other tools based on their own needs. We believe that \toolname will be used as a foundation for future research on dynamic analysis.

\emph{Visual Output.}
\toolname uses the unique state values extracted in the previous step to produce a user-friendly user interface (UI), which we call the \textbf{augmented diff}. This UI adds runtime information to the typical code diff view, e.g. that of GitHub. This new graphical component shows the first unique relevant state values for both the original and patched versions. We consider the first unique state values as the program state differences for two reasons. First, there are usually many unique state values, which is overwhelming for developers. 
Second, the literature has shown that the first point of execution divergence is the most relevant for code reviewers \cite{zhang2019inflection}.

\autoref{fig:UI_Example} shows an example of an augmented diff for a code diff for the Math-20 bug. This generated UI clearly indicates that after changing how variable \texttt{j} is computed in line 1135, the variable is assigned a new value in the patched version. As shown in \autoref{fig:UI_Example}, a code reviewer directly sees that the execution of test \texttt{testMathpbx02} reveals that \texttt{j=24} only occurs in the patched version and \texttt{j=27} only occurs in the original version. By displaying those concrete runtime differences caused by the code change, \toolname helps code reviewers to make an informed decision about the appropriateness of a code diff.

\subsection{Implementation}


\revision{\toolname is an advanced tool made of 5,800 sophisticated lines of Java code, built using Java 11 and relying on ASM and Byte Buddy for instrumentation}. All of the tool's source code is publicly available\footnote{\url{https://github.com/ASSERT-KTH/collector-sahab}}.

\section{Experimental Design}
\label{sec:experimental-design}

\subsection{Research Questions}

\newcommand\rqone{What is the effectiveness of \toolname to detect runtime differences in bug fixing code diffs?}

\newcommand\rqtwo{To what extent does \toolname perform on augmenting diffs from real-world projects?}

\newcommand\rqthree{\revision{What is the quality of the diffs augmented with runtime differences generated by \toolname?}}

\newcommand\rqfour{How do software developers assess \toolname's augmented diffs?}

In this paper, we study the following research questions.

\begin{itemize}
    \item RQ1 (effectiveness): \rqone \ We run \toolname on a large dataset of bug fixing APR code diffs from the DRR \cite{Ye2021EMSE} dataset. We compare \toolname effectiveness with \ddiff \cite{kanda2022didiffff} as baseline.
    
    \item RQ2 (applicability): \rqtwo \ We assess \toolname applicability on real-world commits. This improves the external validity of our evaluation beyond the projects of DRR.
    
    \item RQ3 (quality): \rqthree \ \revision{To assess the quality of \toolname reports for code reviewers, we conduct a manual analysis to evaluate their quality with respect to two criteria: correctness and understandability. We also manually check the  accuracy of \toolname in excluding random runtime differences.} This analysis is performed on the real-world commits also used in the RQ2 experiment.

    \item \revision{RQ4 (user study): \rqfour \ We conduct a user study and ask developers to rate \toolname augmented diffs in terms of usefulness, understandability, and novelty. To compare \toolname against the state of the art, we also ask the participants to give a score to \ddiff outputs. In this study, we use the real-world commits collected for the RQ2 and RQ3 experiment.}
\end{itemize}

\subsection{Datasets}

We use two different datasets for our experiments: \ds and \rqthreeds.

\ds is used for assessing the effectiveness of our tool (RQ1). This dataset contains a subset of code diffs from Ye et al.'s previous research \cite{Ye2021EMSE}. \revision{Ye et al.'s DRR is a collection of 638 code diffs generated by 15 program repair tools for bugs in the Defects4J dataset \cite{just2014defects4j}.} Defects4J includes bugs from five different projects: ``jfreechart'', ``closure-compiler'', ``commons-lang'', ``commons-math'', and ``joda-time''. In \ds, we select DRR code diffs that meet two conditions. First, the project should be a maven project with a ``pom.xml''. \revision{This means we can successfully build both the original and code diffs versions by running a ``mvn compile''.}
Second, the code diff changes should all appear in one single java method.
We focus on single-method code diffs to make sure there is a single, well-scoped behavioral change. 
This means the code diffs that change multiple files, multiple methods, class declarations, method declarations, or imports are ignored.
The result of these two filters is that \ds contains \dssize code diffs from two projects: ``commons-math'' and ``joda-time'', appropriate for our experiments.

\revision{Our second dataset is \rqthreeds, which is used to study \toolname applicability in the field, on a diverse set of complex real-world projects (RQ2 \& RQ3 \& RQ4).} \rqthreeds contains human-made code diffs~\footnote{We consider each commit in GitHub as a code diffs.} from GitHub repositories. To create this dataset, we first start by filtering repositories on GitHub. We select public Java repositories that are not archived and have a Codecov or Coveralls configuration file at the root of the repository. We consider projects using Codecov or Coveralls as  they tend to be well-tested, which makes them appropriate for our experiments. To make sure the chosen repositories are well-established we select repositories that have 50 stars or more. These filters return a list of 158 repositories. Then, we scan these repositories for code diffs, excluding merge code diffs, from 2022-08-13 to 2022-09-12 which change exactly one file that meets two requirements. First, the changed file is a Java file, and second, the whole change lies inside a single method, to be consistent with RQ1. We perform manual analysis on each code diff to eliminate those where the code diff is not covered by a test, the project fails to compile, or we fail to execute the tests that cover the change. The resulting \rqthreeds dataset contains \revision{50} code diffs from 10 GitHub repositories as shown in \autoref{tab:repositories}.

In \autoref{tab:repositories}, we give the total number of commits, modules, and lines of code (KLOC) per considered project to give a sense of their complexity. As shown in the table, the median number of code diffs, modules, and lines of code are 6,069, 8, and 175K. This indicates that the projects in \rqthreeds are established repositories with real-world complexity. These statistics confirm that the \rqthreeds code diffs are selected from complex, real-world repositories. 
If \toolname works on them, it is a good indicator of the quality of our approach and its prototype.

\begin{table}[t]
\centering
\footnotesize
\caption{Descriptive statistics of open-source projects included in benchmark \rqthreeds.}
\label{tab:repositories}
\begin{tabular}{@{}l r r r@{}}
\toprule
	Repository & \#commits & \#modules & KLOC \\
	\midrule
	\href{https://github.com/codingapi/springboot-framework/}{codingapi/springboot-framework} & 178 & 5 & 3 \\
	\href{https://github.com/nosan/embedded-cassandra/}{nosan/embedded-cassandra} & 2,149 & 1 & 12 \\
	\href{https://github.com/infinitest/infinitest/}{infinitest/infinitest} & 489 & 7 & 25 \\
	\href{https://github.com/jhipster/jhipster-lite/}{jhipster/jhipster-lite} & 6,413 & 1 & 90 \\
	\href{https://github.com/schemacrawler/SchemaCrawler/}{schemacrawler/SchemaCrawler} & 8,198 & 23 & 148 \\
	\href{https://github.com/alibaba/nacos/}{alibaba/nacos} & 4,224 & 18 & 203 \\
	\href{https://github.com/apache/dubbo/}{apache/dubbo} & 5,725 & 27 & 229 \\
	\href{https://github.com/cdk/cdk/}{cdk/cdk} & 17,263 & 9 & 502 \\
	\href{https://github.com/apache/iotdb/}{apache/iotdb} & 7,364 & 43 & 591 \\
	\href{https://github.com/JMRI/JMRI/}{JMRI/JMRI} & 74,997 & 1 & 4,741 \\
    \midrule
    Min & 178 & 1 & 3 \\
    Median & 6,069 & 8 & 175 \\
    Max & 74,997 & 43 & 4,741 \\
	\bottomrule
\end{tabular}
\end{table}

\subsection{Baseline}
\label{sec:baseline}
There are very few runtime differencing tools that are available. In our work, we focus on Java, and there is one single open-source runtime differencing tool for Java, \ddiff (pronounced di-di-fu-fu) \cite{kanda2022didiffff}.

Consequently, in our experimental evaluation, we consider \ddiff as the baseline. \ddiff runs two versions of a program and illustrates the difference between the values assigned to program variables in a web interface. \ddiff uses \selogger \cite{shimari2019near} to record all the values each variable of a program holds during an execution. Using these values, \ddiff creates two lists of values for each variable access that is not part of the code change. The first list represents the values assigned to the variable at that point of access in the original version. The second list represents values assigned to the variable in the patched version of the program. Finally, \ddiff compares the two lists of values for each variable access. The result of this comparison is presented to code reviewers in form of a graphical user interface (GUI). This GUI shows the two lists of values assigned to a variable and highlights if there is a difference between these two lists. \revision{An example output of \ddiff can be seen at \url{https://bit.ly/3N7TKEb}}.

\ddiff reports two types of runtime differences: 1) the difference in the content of the list of values assigned to primitive variables, and 2) the difference in the length of the list of values assigned to non-primitive variables. Regarding non-primitive variables, \ddiff only counts how many values are assigned to them, it does not consider what is inside those values. That is because \ddiff does not look deep into the objects referred by non-primitive variables, which makes it significantly different from \toolname.


\subsection{RQ1: Effectiveness on Benchmark}
\label{sec:rq1_protocol}

To answer \textbf{RQ1}, we run \toolname on all \dssize code diffs in \ds with four different state depths: depth=0,1,2,3. This maximum depth of three has been found empirically, because collecting data for higher depths requires resources beyond our experimental setup without improving effectiveness.

The key metric is the number of code diffs for which \toolname augments the diff. \revision{We split the remaining code diffs into three groups and count the number of code diffs in each group. Code diffs for which no unique state value is detected, code diffs that cause a memory failure, and code diffs that make \toolname exceed the time limit.}

We also run \ddiff on the \ds code diffs. We count the number of code diffs for which \ddiff detects a runtime difference contrasting it with the number of code diffs for which \toolname augments the diff.

We perform our experiment on a machine with sixteen Intel(R) Core(TM) i9-10980XE processors, each running at 3.00GHz and having 18 cores, and eight 16GB RAMs each of type DDR4 with 3600 MT/s clock speed.

\subsection{RQ2: Applicability in the Field}
\label{sec:rq2_protocol}
To answer \textbf{RQ2}, we run \toolname on all code diffs in \rqthreeds. Then, we compute the set of code diffs for which \toolname successfully runs and lists all the unique state values (if any). These are the \emph{successful code diffs}. Next, we get all the repositories corresponding to these code diffs and call the set of these repositories as \emph{successful repositories}.

Finally, we compute the ratio $AC=\frac{|successfull\_code\_diffs|}{|all\_code\_diffs|}$, which represents the fraction of code diffs on which \toolname works successfully and accurately. A high value of $AC$ and a large set of successful repositories indicate how applicable our tool is on real-world projects and, from a scientific perspective, measure the external validity of our results.

\revision{\subsection{RQ3: Manual Analysis on Augmentation Quality}}
\label{sec:rq3_protocol}

In answer to \textbf{RQ3}, we conduct a detailed manual analysis of the outputs of running \toolname on \rqthreeds code diffs. \revision{In this analysis, we consider all the successful code diffs in \rqthreeds.} This consists of the code diffs for which \toolname detects a unique state value, and code diffs for which \toolname does not detect any unique state values. Note that all code diffs in \rqthreeds are covered by a test, but the covering test is not required to fail on the original version and pass on the patched version. This means the test may not reveal any runtime difference.

\revision{For code diffs with no detected unique state value, we manually verify that the code change does not entail any unique state value in neither the original nor the patched version. For each code diff with a detected unique state value, we study the quality of \toolname augmented diff from three aspects: correctness, understandability, and randomness exclusion accuracy of the detected runtime difference.} In case of negative results for any of the three mentioned criteria, we investigate the reasons behind them.

To assess correctness, we manually run the covering test line by line in a debugger twice, first on the original and second on the patched versions of the program under study. We ensure the first unique relevant state value reported by \toolname matches what we see in the debugger value explorer.

Regarding understandability, we analyze whether a code reviewer could understand the reported runtime behavioral change in a reasonable amount of time (typically 5-30 minutes, per typical code review practices at the function level).

\revision{Finally, we evaluate the randomness exclusion accuracy of \toolname to see if its \texttt{random-exclusion} configuration (as defined in \autoref{sec:program_state_diffing}) works accurately and effectively. We perform this evaluation in two steps as follows.}

\revision{First, we divide the commits into two groups. Group 1 contains commits where the unique state value is a non-random value and Group 2 contains commits wherein the unique state value is reported only because of differences in random value. We filter Group 2 commits by setting \texttt{random-exclusion=false} and then running \collectorsahab again on these commits with \texttt{random-exclusion=true}. If the unique state value reported disappears, we consider such commits under Group 2. If after changing \texttt{random-exclusion} a unique state value is still reported, we consider the commit under Group 1.}

\revision{In the second step, we manually perform two checks. In the first check, we investigate if the unique state values reported for commits in Group 2 are  actually random and no other non-random unique state value exists. If our manual analysis confirms this, it means \toolname has performed accurately by not identifying any unique state value for these commits when \texttt{random-exclusion=true}. Second, we check if the unique state values reported for commits in Group 1 are not random, when \texttt{random-exclusion=true}. Again, if our manual analysis confirms this, it indicates the accuracy of randomness exclusion by \toolname. A high accuracy of randomness exclusion on the commits means that the \texttt{random-exclusion} configuration of \toolname is an accurate and highly applicable tool for excluding runtime differences that are caused by flaky behavior of programs.}

This analysis is conducted by two of the authors, both experts in Java programming. In case of any discrepancies between the participants, they meet and discuss to resolve them and reach a conclusion.

\revision{\subsection{RQ4: User Study}
\label{sec:rq4_protocol}
To answer \textbf{RQ4}, we perform a user study. In this study, we ask experienced developers to comment on the quality of \toolname and \ddiff reports. For this experiment, we have two groups of participants: a group of four participants working in industry, and a group of four participants studying at our university. All the participants are experienced in programming with Java language and creating and merging GitHub pull requests.}

\revision{For each participant, the experiment is performed in three steps as follows. First, in a live presentation, we use some examples to introduce the relevant concepts of runtime differencing and the two tools (\toolname and \ddiff) to the participant. Second, we present a selected set of commits and their augmentations with both tools. For each commit, the participant gives a score between 1 and 5 to each augmented code diff in terms of \textbf{usefulness}, \textbf{clarity}, and \textbf{novelty}. Usefulness means whether the report generated by \toolname helps in understanding the changes made to the program. Clarity means how comprehensible the generated report is. Novelty measures how novel is the information provided by \toolname in the sense that it cannot be obtained by looking at the plain code diff. In the third and final step, we hold a short discussion with the participant when they have finished rating the augmented code diffs. In the final discussion, we ask for an overall feedback regarding code diff augmentation with runtime data, how the tools compare against each other, and potential improvements in each tool.}

\revision{To select the commits that are shown to participants, we consider the \rqthreeds commits that cause a runtime difference per our manual analysis in RQ3. We exclude three types of commits from the RQ3 experiment. First, the commits that suffer from random runtime differences. Second, if there are multiple commits changing the same line of a program, we only show one of them to a participant. 
Third, we also exclude commits that change a test method.
After applying these filters, we select a total of 8 commits. We split these commits into two groups (industry and students) so that each group gets 4 commits to analyze. This is is an appropriate number as it fits well within the one-hour time limit that we set for each participant.}

\revision{
In answer to each question, the participants give a score between 1 and 5 to the augmented code diff. To have an overall assessment of the given scores, we first compute the average score a participant has given to the augmented diffs of each tool in answer to each question. Then, we calculate the median score given by participants in each group (industry/students) to obtain the overall assessment in relation to each criterion.}

\revision{Note that for the commits for which only one of the tools detects a runtime difference, the scores cannot be compared between the two tools. Therefore, we consider these commits separately.}

\revision{In our final live discussion with participants, we ask them three questions. First, if they think augmenting code diffs with runtime data can be a useful practice. Second, we ask them which tool they prefer overall. Finally, we ask whether they have any specific suggestion for improving each tool. The answers given by participants help us to see if this path of code diff augmentation is promising and what are the best ways to improve the state of the art.}

\section{Experimental Results}
\label{sec:experimental-results}

\subsection{\toolname Effectiveness (RQ1)}
\label{sec:rq1_res}


\begin{table*}[t]
\centering
\caption{\revision{Results of running \toolname on \dssize code diffs in \ds.}}
\label{tab:rq1_results}
\begin{threeparttable}[b]
\begin{tabular}{@{}l r | r r r@{}}
    \toprule
	Diff Tool & Augmented Diff & No Diff Detected & Memory Failure & Time Limit \\
	\midrule
	\ddiff & 377 (64.5\%) & 207 (35.5\%) & -- & -- \\ 
    \toolname (depth=0) & 543 (93.0\%) & 32 (4.9\%) & 9 (1.5\%) & 0 (0.0\%) \\
    \toolname (depth=1) & 555 (95.0\%) & 18 (3.2\%) & 9 (1.5\%) & 2 (0.3\%) \\
    \toolname (depth=2) & 554 (94.8\%) & 16 (2.8\%) & 13 (2.2\%) & 1 (0.2\%) \\
    \toolname (depth=3) & 544 (93.1\%) & 22 (3.9\%) & 18 (3.0\%) & 0 (0.0\%) \\
	\bottomrule
\end{tabular}
\begin{tablenotes}[flushleft]
\end{tablenotes}
\end{threeparttable}
\end{table*}

\autoref{tab:rq1_results} shows the results of running \toolname on \dssize code diffs from our benchmark \ds. In each row, ``Diff Tool'' shows the used tool and its configuration. More specifically, we run \toolname with four different depths from zero to three and \ddiff on the dataset (see \autoref{sec:terminology} and \autoref{sec:trace_collection}). The second column ``Augmented Diff'' indicates the number of code diffs for which a runtime difference is detected and the diff is augmented with a unique state value. Each of the remaining three columns explains the failure modes.  \revision{``No Diff Detected'' indicates the number of code diffs for which no unique state value in either of the versions is found (see \autoref{sec:output}).} Finally, ``Memory Failure'' and ``Time Limit'' show the number of code diffs on which \toolname faces memory and time limit during execution, respectively.

\revision{Consider ``\toolname (depth=1)'' in \autoref{tab:rq1_results} as an example. The table shows that in this case, \toolname augments the diff with a detected unique state value for 95\% (555/\dssize) code diffs. For most code diffs, \toolname is reliable to identify an execution difference between two versions of a program. With depth set to 1, \toolname faces the memory limit for 1.5\% (9/\dssize) of the code diffs and fails due to exceeding the time limit only for 0.3\% (2/\dssize) of the code diffs, which represents a tiny minority. For the remaining 3.2\% (18/\dssize) of the diffs, \toolname runs successfully but does not detect any unique state value in either of the versions. This shows for most of the code diffs the differentiating test triggers a unique state value that is identified and reported by \toolname.}

\revision{As shown in \autoref{tab:rq1_results}, \toolname outperforms the state-of-the-art tool \ddiff. 
The number of code diffs for which we have a runtime difference is higher for all depths, from 93\% (543/\dssize) to 95\% (555/\dssize) for different depths, while it is 64.5\% (377/\dssize) for \ddiff. \toolname detects and reports a runtime difference in notably more cases than \ddiff.}

Next, we manually analyze the code diffs for which \toolname or \ddiff fail to generate a diff report.
Our analysis indicates that there are three reasons why \toolname outperforms \ddiff in terms of identifying runtime differences. \revision{First, \ddiff only considers differences between values of primitive types, while \toolname also considers differences between the value of non-primitive variables that refer to objects.} Second, \ddiff only considers differences between local variable values, while \toolname also detects the difference between the values returned by methods (inside a return statement without an explicit variable). The third and last reason that \toolname outperforms \ddiff is related to \ddiff not logging the variable values inside inner classes. The latter case is present in our dataset. These three reasons explain why \toolname is able to detect more runtime differences compared to \ddiff.

\revision{The column ``No Diff Detected'' shows the number of code diffs for which \toolname does not find a unique state vlue. As shown in \autoref{tab:rq1_results}, generally, the number of these cases decreases when depth increases: the number of code diffs with no diff detected is 32 for depth=0; this number decreases to 16 for depth=2. This indicates that \toolname's power to look deep inside non-primitive objects adds to its effectiveness in detecting runtime differences. There is an exception for depth=3, where the number of code diffs without a detected diff increases to 22. Per our detailed manual analysis of the logs, the reason behind this is a problem with one of the libraries that \toolname uses. This library \footnote{\url{https://github.com/FasterXML/jackson}} fails serialize very large Java object into JSON format. As the size of trace object collected for depth=3 is as large as 10GB in some cases, this leads to \toolname failure to detect a runtime difference. This shows increasing the depth has its own cost and we should carefully select an optimal depth for configuring \toolname in a given setup.}

In the current version, \toolname stores all the collected data in RAM during execution and prints the whole trace of the program under test at the end. As the entire trace may be very large, it causes memory failure in some cases. The ``Memory Failure'' column of \autoref{sec:rq1_res} shows that facing a memory limit during \toolname execution is rare. \revision{When the depth is set to zero or one, we face memory limit issues only for 1.5\% (9/\dssize) code diffs. When depth is set to two or three, the memory limit causes an issue for 2.2\% (7/\dssize) and 3.0\% (18/\dssize) of code diffs, respectively.} There is room for further improvement in this regard, as memory failures can be partially avoided by printing the execution trace step-by-step. This means there is a significant potential to improve \toolname memory usage.

\revision{\toolname puts breakpoints at all matched lines \autoref{sec:breakpoint_identification}, which means the instrumenter collects the program state at many places. This adds an overhead during the runtime and makes the execution take longer than when the code is not instrumented. With respect to time, the ``Time Limit'' column shows that we rarely face cases where \toolname fails due to time issues. We exceed the time limit only with depth=1 and depth=2 and only for 0.3\% (2/\dssize) and 0.2\% (1/\dssize) of the code diffs. This means \toolname is fast enough to finish its analysis on most code diffs. We note that \toolname exceeds the time limit for more cases when depth=1, compared to when depth=2 and depth=3. This happens while increasing the depth should make the tool take more time to collect the trace and compute the diff. The reason is that when the depth is increased, we may face memory failure or the library issue mentioned above even before exceeding the time limit. Therefore, the ``Time Limit'' case turns into ``Memory Failure'' or ``No Diff Detected'' in such cases.}

\revision{Overall, we notice that \toolname performs the best when the depth is set to one by augmenting the diff for 95\% (555/\dssize) of code diffs.} Depth=1 outperforms other depths as it provides an accurate balance between looking at details (better than depth=0) and collecting too much data (better than depth=2,3).

\begin{mdframed}\noindent
    \textbf{Answer to RQ1: \rqone} \\
    Our experiment on benchmark \ds shows that \toolname is effective at detecting runtime differences. \revision{\toolname augments diffs with runtime information for 95\% (555/\dssize) of the code diffs in our benchmark, outperforming \ddiff, which detects a runtime difference in only 64.5\% (377/\dssize) of the code diffs.} It is clear that \toolname enhances the state of the art of runtime differencing.
\end{mdframed}

\subsection{\toolname Applicability (RQ2)}

\begin{table}[t]
\centering
\caption{Results of running \toolname on \rqthreeds.}
\label{tab:field_results}
\begin{tabular}{@{}l r r r | r@{}}
\toprule
    Repository & |ALL| & Success & AC & Failure \\
    \midrule
    \href{https://github.com/JMRI/JMRI/}{JMRI/JMRI} & 13 & 12 & 0.92 & 1 \\
    \href{https://github.com/alibaba/nacos/}{alibaba/nacos} & 2 & 2 & 1.00 & 0 \\
    \href{https://github.com/apache/dubbo/}{apache/dubbo} & 4 & 4 & 1.00 & 0 \\
    \href{https://github.com/apache/iotdb/}{apache/iotdb} & 8 & 7 & 0.88 & 1 \\
    \href{https://github.com/cdk/cdk/}{cdk/cdk} & 5 & 5 & 1.00 & 0 \\
    \href{https://github.com/codingapi/springboot-framework/}{codingapi/springboot-framework} & 1 & 1 & 1.00 & 0 \\
    \href{https://github.com/infinitest/infinitest/}{infinitest/infinitest} & 1 & 1 & 1.00 & 0 \\
    \href{https://github.com/jhipster/jhipster-lite/}{jhipster/jhipster-lite} & 7 & 7 & 1.00 & 0 \\
    \href{https://github.com/nosan/embedded-cassandra/}{nosan/embedded-cassandra} & 1 & 1 & 1.00 & 0 \\
    \href{https://github.com/schemacrawler/SchemaCrawler/}{schemacrawler/SchemaCrawler} & 8 & 8 & 1.00 & 0 \\
    \midrule
    Total & 50 & 48 & 0.96 & 2 \\
	\bottomrule
\end{tabular}
\end{table}

In our study of \toolname applicability, we run \toolname on code diffs from benchmark \rqthreeds. The results of this experiment are shown in \autoref{tab:field_results}. In this table, the left column indicates the repository id. The number of all code diffs from each real-world repo is shown in the ``|ALL|'' column, per our selection criteria. As defined in \autoref{sec:rq2_protocol}, ``Success'' represents the number of all code diffs successfully processed by \toolname. ``AC'' is the ratio of code diffs successfully processed by \toolname and finally, ``Failure'' indicates the number of code diffs that \toolname fails to process. 

\revision{Consider the repository jhipster/jhipster-lite as an example (eighth row). \rqthreeds consists of 7 code diffs from this repository. \toolname successfully performs on all 7 code diffs, giving $AC$ equal to 1 (7/7).}

In total, our selection criteria yield 50 code diffs from real-world complex projects. They are sampled from a set of ten diverse open-source projects. \revision{\toolname successfully analyzes 48 code diffs out of a total of 50.} For all repositories, we have at least one successfully processed code diff. This indicates that \toolname is successful in handling diverse Java projects from \rqthreeds. \revision{With 50 total code diffs and 48 successfully analyzed ones, the corresponding value of AC is 0.96 (48/50)}. Notably, the median AC is 1, because \toolname handles all code diffs for 6 projects. Such a high value of AC attests the real applicability of \toolname.

\revision{Finally, as shown in the ``Failure'' column, there are 2 commits where \toolname fails to process the code diff to collect a runtime diff.} \revision{Both of them are due to out-of-memory exception.} This can be attributed to the huge number of states collected. In these cases, either the tool runs out of heap space while generating the execution trace (step `trace collection') or while iterating through states to find a diff (step `program state differencing'). It is known that fine-grain monitoring is very memory intensive  \cite{barr2014tardis}.

\begin{mdframed}\noindent
    \textbf{Answer to RQ2: \rqtwo} \\
    To analyze applicability in the wild, we build a well-formed benchmark of 50 code diffs from 10 complex open-source Java repositories. \revision{\toolname successfully runs on 48 code diffs, which yields a high success ratio of \revision{0.96}.} \toolname is able to handle real-world Java projects, incl. large ones with multi-module build systems.
\end{mdframed}

\revision{\subsection{\toolname Augmentation Quality (RQ3)}}
\label{sec:rq3_results}

\revision{Per the RQ3 protocol, we run \toolname with two configurations (\texttt{random-exclusion=true,false}) and use a Java debugger to analyze the \rqthreeds augmented code diffs.}

\revision{First, we analyze the 18 code diffs for which \toolname does not detect a unique state value even with \texttt{random-exclusion=false}.} Our manual analysis through the debugger verifies what \toolname states about these code diffs: they do not entail any unique state value. This indicates the correctness of \toolname results on this set of code diffs.

\begin{table*}[t]
\centering
\footnotesize
\caption{\revision{Manual analysis of the quality of \toolname's augmented diffs on real-world code changes.}}
\label{tab:report_quality}
\begin{tabular}{l l l c c}
\toprule
	& ID & \textbf{Code diff} & Correctness & Understandability \\
	\midrule
    \parbox[t]{2mm}{\multirow{16}{*}{\rotatebox[origin=c]{90}{\textbf{\texttt{rand-exc=true}}}}}
	& 1 & \href{https://assert-kth.github.io/collector-sahab-experiments/rq2-random-excluded-res/SchemaCrawler_07c7368}{SchemaCrawler@07c7368} & \cmark & \xmark \\
	& 2 & \href{https://assert-kth.github.io/collector-sahab-experiments/rq2-random-excluded-res/SchemaCrawler_55a55de}{SchemaCrawler@55a55de} & \cmark & \cmark \\
	& 3 & \href{https://assert-kth.github.io/collector-sahab-experiments/rq2-random-excluded-res/SchemaCrawler_f950519}{SchemaCrawler@f950519} & \cmark & \cmark \\
	& 4 & \href{https://assert-kth.github.io/collector-sahab-experiments/rq2-random-excluded-res/JMRI_0b53ea8}{JMRI@0b53ea8} & \cmark & \cmark \\
	& 5 &\href{https://assert-kth.github.io/collector-sahab-experiments/rq2-random-excluded-res/JMRI_4a30042}{JMRI@4a30042} & \cmark & \cmark \\
	& 6 & \href{https://assert-kth.github.io/collector-sahab-experiments/rq2-random-excluded-res/iotdb_5faa9da}{iotdb@5faa9da} & \cmark & \xmark \\
    & 7 & \href{https://assert-kth.github.io/collector-sahab-experiments/rq2-random-excluded-res/iotdb_a3559e5}{iotdb@a3559e5} & \cmark & \cmark \\
	& 8 & \href{https://assert-kth.github.io/collector-sahab-experiments/rq2-random-excluded-res/iotdb_e5e4f17}{iotdb@e5e4f17} & \cmark & \cmark \\
	& 9 & \href{https://assert-kth.github.io/collector-sahab-experiments/rq2-random-excluded-res/springboot-framework_70b0d12}{springboot-framework@70b0d12} & \cmark & \cmark \\
    & 10 & \href{https://assert-kth.github.io/collector-sahab-experiments/rq2-random-excluded-res/jhipster-lite_510a975}{jhipster-lite@510a975} & \cmark & \cmark \\
	& 11 & \href{https://assert-kth.github.io/collector-sahab-experiments/rq2-random-excluded-res/jhipster-lite_5d2a6b8}{jhipster-lite@5d2a6b8} & \cmark & \cmark \\
	& 12 & \href{https://assert-kth.github.io/collector-sahab-experiments/rq2-random-excluded-res/jhipster-lite_69a7d5a}{jhipster-lite@69a7d5a} & \cmark & \cmark \\
	& 13 & \href{https://assert-kth.github.io/collector-sahab-experiments/rq2-random-excluded-res/jhipster-lite_9162791}{jhipster-lite@9162791} & \cmark & \cmark \\
	& 14 & \href{https://assert-kth.github.io/collector-sahab-experiments/rq2-random-excluded-res/jhipster-lite_e2fc61f}{jhipster-lite@e2fc61f} & \cmark & \cmark \\
	& 15 & \href{https://assert-kth.github.io/collector-sahab-experiments/rq2-random-excluded-res/jhipster-lite_f5b592a}{jhipster-lite@f5b592a} & \cmark & \cmark \\
    & 16 & \href{https://assert-kth.github.io/collector-sahab-experiments/rq2-random-excluded-res/cdk_d500be0}{cdk@d500be0} & \cmark & \cmark \\
    & 
    17 & \href{https://assert-kth.github.io/collector-sahab-experiments/rq2-random-excluded-res/dubbo_29e4e42.html}{dubbo@29e4e42} & \cmark & \cmark \\
    \midrule
    \parbox[t]{2mm}{\multirow{11}{*}{\rotatebox[origin=c]{90}{\textbf{\texttt{rand-exc=false}}}}}
    & 18 & \href{https://assert-kth.github.io/collector-sahab-experiments/rq2-random-included-res/embedded-cassandra_dcc5818d80489d3ebc2ed4265d673bc05af08551_1.html}{embedded-cassandra@dcc5818} & \cmark & \cmark \\
    & 19 & \href{https://assert-kth.github.io/collector-sahab-experiments/rq2-random-included-res/SchemaCrawler_5955cd0506697c04a6dd60f742a4ca53c2c5d9ae_3.html}{SchemaCrawler@5955cd0} & \cmark & \cmark \\
    & 20 & \href{https://assert-kth.github.io/collector-sahab-experiments/rq2-random-included-res/SchemaCrawler_c06131a25e1dc30e8d2d00d49b6ad06c4fa0c126_3.html}{SchemaCrawler@c06131a} & \cmark & \cmark \\
    & 21 & \href{https://assert-kth.github.io/collector-sahab-experiments/rq2-random-included-res/SchemaCrawler_451a5a5a60c76acb7303ffa734c0fedc6ad7de19_3.html}{SchemaCrawler@451a5a5} & \cmark & \cmark \\
    & 22 & \href{https://assert-kth.github.io/collector-sahab-experiments/rq2-random-included-res/JMRI_3e2427222f6e6822972d54c5a472d493733a01e3_3_random.html}{JMRI@3e24272} & \cmark & \cmark \\
    & 23 & \href{https://assert-kth.github.io/collector-sahab-experiments/rq2-random-included-res/JMRI_d21837e89e76e733333646f7dc02746803164aad_3_random.html}{JMRI@d21837e} & \cmark & \cmark \\
	& 24 &\href{https://assert-kth.github.io/collector-sahab-experiments/rq2-random-included-res/dubbo_bb7f942c5dc1649ff8e0be939639280e44c731a7_0.html}{dubbo@bb7f942} & \cmark & \cmark \\
	& 25 & \href{https://assert-kth.github.io/collector-sahab-experiments/rq2-random-included-res/iotdb_35541f457afa7e3075893c633bdb1ea8765a399c_3.html}{iotdb@35541f4} & \cmark & \cmark \\
    & 26 & \href{https://assert-kth.github.io/collector-sahab-experiments/rq2-random-included-res/iotdb_94657cf836ea6407cfd374f36324333e8194b919_3.html}{iotdb@94657cf} & \cmark & \cmark \\
	& 27 & \href{https://assert-kth.github.io/collector-sahab-experiments/rq2-random-included-res/iotdb_e5e4f17f703fa7c8e7623d0a8d3120acb0e89fa6_3.html}{iotdb@e5e4f17} & \cmark & \cmark \\
	& 28 & \href{https://assert-kth.github.io/collector-sahab-experiments/rq2-random-included-res/jhipster-lite_2b3879966bb1a7a54e8fa90a207e81f37fd1332f_3/1.html}{jhipster-lite@2b38799} & \cmark & \cmark \\
    & 
    29 & \href{https://assert-kth.github.io/collector-sahab-experiments/rq2-random-included-res/dubbo_6b095f1c68b546c9f0ce72dd91b9c1d2792c60be.html}{dubbo@6b095f1} & \cmark & \cmark \\
    & 
    30 & \href{https://assert-kth.github.io/collector-sahab-experiments/rq2-random-included-res/nacos_0cf9c24086ac904e350fc53b4b9f44e0f439e318.html}{nacos@0cf9c24} & \cmark & \cmark \\
    \midrule
	& \multicolumn{2}{l}{Total} & 100\% (30/30) & 93\% (28/30) \\ 
    \bottomrule
\end{tabular}
\end{table*}

\begin{figure*}
\begin{center}
\includegraphics[width=\textwidth]{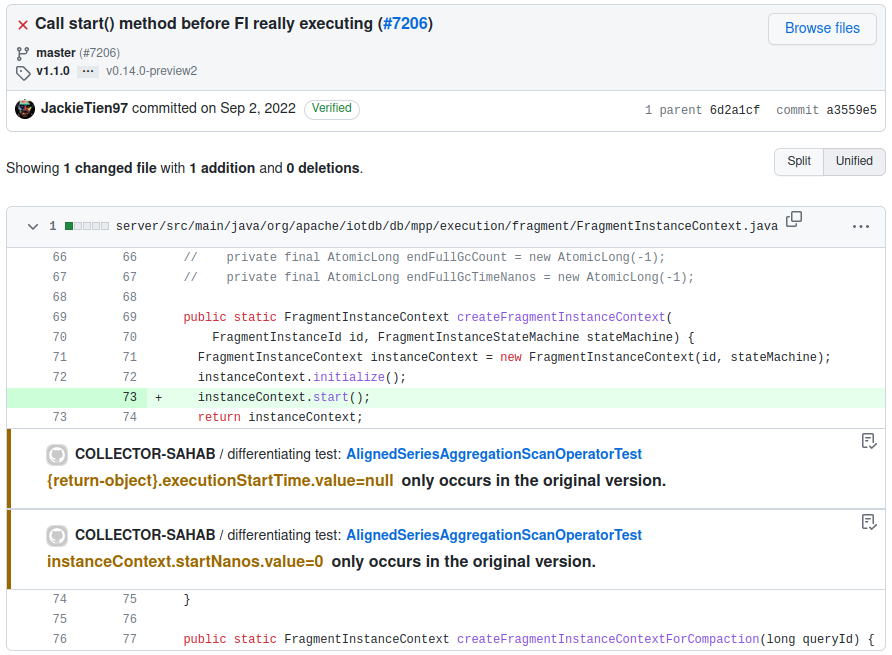}
\caption{Augmented diff for apache/iotdb@a3559e5, accurately pointing to the first unique relevant state value in the runtime data.}
\label{fig:rq2_sc}
\end{center}
\end{figure*}

\revision{More importantly, there are 30 code diffs for which \toolname detects a unique state value and generates an augmented diff. The list of these code diffs is presented in \autoref{tab:report_quality}. As shown in the table, for 17 of the code diffs, \toolname outputs a unique state value with \texttt{random-exclusion=true}. For the remaining 13 code diffs, \toolname detects a unique state value only with \texttt{random-exclusion=false}. Therefore, \toolname labels 40\% (13/30) of the considered code diffs as causing random runtime differences. This indicates the notable importance of the random value exclusion feature on real-world commits.}

\revision{\autoref{tab:report_quality} also presents the results of our manual analysis on the correctness, understandability, and randomness exclusion accuracy of the runtime difference detected for the code diffs. We define these metrics as follows. Column ``Correctness'' shows if \toolname's augmented diff matches the difference we witness by debugging the code. The ``Understandability'' column shows if it is possible to understand the augmented diff in a reasonable amount of time.}

\revision{Let's take the code diff \#7 as an example, from project apache/iotdb. \toolname generates the augmented diff for this code diff as shown in \autoref{fig:rq2_sc}. The message of this commit reads as ``Call start() method before FI really executing'' and the plain text code diff also shows \texttt{instanceContext.start()} is called on line 73 of the patched version. However, this does not show how calling \texttt{instanceContext.start()} affects the runtime behavior of the program. Here \toolname comes in and claims that when this method is not called in the original version, \texttt{startNanos.value} of \texttt{instanceContext} is set to 0 and calling this method in the patch changes this value. This gives the code reviewer a very concrete and detailed runtime difference information that cannot be easily obtained just by looking at the textual diff. Our manual debugging of the program verifies the correctness of \toolname's report for this code diff. This reported runtime difference in this augmented diff is also easy to understand.}

\revision{Overall, the execution difference reported by \toolname is correct for all 30 code diffs in our benchmark. Generating a correct augmented diff for 100\% of the code diffs indicates that \toolname is a reliable tool for execution difference detection. Our manual analysis of the randomness exclusion accuracy also shows that \toolname correctly determines whether unique state values are random or not for 100\% of the code diffs. In other words, \toolname is able to correctly exclude unique state values that are flaky or related to file paths as explained in \autoref{sec:program_state_diffing}. This indicates that \toolname's augmented diffs help developers focus on non-random and runtime differences causally related to the change.}

\revision{We also see that the \toolname output for code diff \#6 is not easy to understand.} By carefully analyzing this code diff, we see that the reported runtime difference is hard to understand because the code change is a multi-line change and all changed lines affect the reported unique state value. To understand the augmented diff, we need to know the value of many variables involved in the code diff. Recall that we designed \toolname to only show the first unique relevant state value. This demonstrates the trade-off between conciseness and comprehensiveness of the augmented diff. \revision{To summarize, \toolname's report for the code diff \#6 is not easy to understand because more variable values are necessary for this particular case.}

\begin{figure}
\begin{center}
\includegraphics[width=\columnwidth]{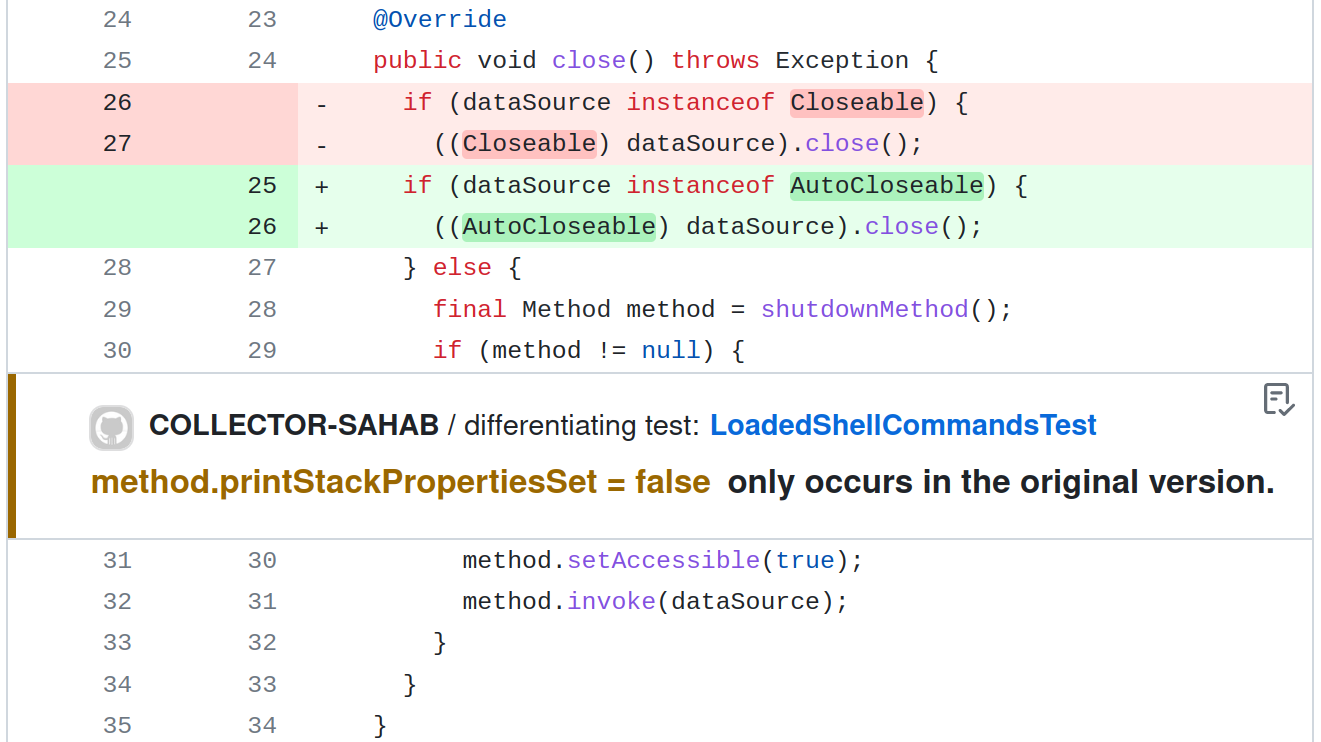}
\caption{The augmented diff for SchemaCrawler@07c7368 shows a runtime diff that is hard to understand, due to reporting a static field value for a null object.}
\label{fig:rq2_ununderstandable_ex}
\end{center}
\end{figure}

\revision{The code diff \#1 is another case where the augmentation by \toolname is not easily understandable. This augmented code diff is shown in \autoref{fig:rq2_ununderstandable_ex}. The detected unique state value is the \texttt{false} value of the \texttt{printStackPropertiesSet} field in the \texttt{method} variable. The issue with this runtime data is that it does not state if \texttt{printStackPropertiesSet} is actually a static field. Consequently, the user may imply that \texttt{method} is referring to a not-null object whose \texttt{printStackPropertiesSet} is set to \texttt{true}, while in fact the value of \texttt{method} is null in this example. Based on this observation, we envision a future improvement in \toolname that adds the type (static/non-static) of the fields to the reported runtime differences.}

\begin{mdframed}\noindent
    \textbf{Answer to RQ3: \rqthree} \\
    \revision{The manual analysis of 30 code diffs in the benchmark \rqthreeds shows that 100\% of augmented diffs are correct. The randomness of the detected unique state values is correctly determined in 100\% as well. For 93\% (28/30) of the code diffs, the reported runtime difference is easily understandable. Overall, based on our experiments, we conclude that \toolname is a reliable and useful tool to understand code changes, beyond purely syntactic code diffs.}
\end{mdframed}

\revision{\subsection{\toolname User Study (RQ4)}}

\begin{table*}[t]
\centering
\footnotesize
\caption{\revision{Commits considered for the RQ4 experiment, their augmenting tool(s), and assigned group.}}
\label{tab:rq4_commits}
\begin{tabular}{l l c c r}
\toprule
	ID & \textbf{Commit} & \toolname & \ddiff & Group \\
	\midrule
	\rowcolor{black!10} C1 & \href{https://github.com/cdk/cdk/commit/d500be0}{cdk@d500be0} & \cmark & \cmark & Industry \\
     C2 & \href{https://github.com/apache/iotdb/commit/5faa9da}{iotdb@5faa9da} & \cmark & \cmark & Industry \\
     \rowcolor{black!10} C3 & \href{https://github.com/schemacrawler/SchemaCrawler/commit/55a55de}{SchemaCrawler@55a55de} & \cmark & \cmark & Industry \\
     C4 & \href{https://github.com/jhipster/jhipster-lite/commit/9162791}{jhipster-lite@9162791} & \cmark & \xmark & Industry \\
     \midrule
     \rowcolor{black!10} C5 & \href{https://github.com/schemacrawler/SchemaCrawler/commit/07c7368}{SchemaCrawler@07c7368} & \cmark & \cmark & Students \\
     C6 & \href{https://github.com/apache/iotdb/commit/e5e4f17}{iotdb@e5e4f17} & \cmark & \cmark & Students \\
     \rowcolor{black!10} C7 & \href{https://github.com/schemacrawler/SchemaCrawler/commit/f950519}{SchemaCrawler@f950519} & \cmark & \cmark & Students \\
     C8 & \href{https://github.com/jhipster/jhipster-lite/commit/f5b592a}{jhipster-lite@f5b592a} & \cmark & \xmark & Students \\
    \bottomrule
\end{tabular}
\end{table*}

\revision{After applying the filters mentioned in \autoref{sec:rq4_protocol}, we end up with eight commits shown in \autoref{tab:rq4_commits}. It shows, per commit, the tools that detects a runtime difference. Note that \toolname detects a runtime difference for all commits, while \ddiff fails to do this for C4 and C8. The last column of the table shows the group that the commit is assigned to. As explained in \autoref{sec:rq4_protocol}, the commits are split into two groups, one group for participants from industry and one group for students. This gives us four commits for each group.}

\begin{table*}[t]
\centering
\small
\caption{\revision{Scores given by each participant to augmented diffs generated by \toolname and \ddiff. ``CS'' and ``DD'' represent \toolname and \ddiff, respectively.}}
\label{tab:user_study_res}
\begin{tabular}{l l | r r | r r | r r | r}
\toprule
    & & \multicolumn{2}{c}{\footnotesize Usefulness} & \multicolumn{2}{c}{\footnotesize Clarity} & \multicolumn{2}{c}{\footnotesize Novelty} & \\
    & \textbf{Particpant ID} & CS & DD & CS & DD & CS & DD & \textbf{Preference} \\
	\midrule
    	 \parbox[t]{2mm}{\multirow{5}{*}{\rotatebox[origin=c]{90}{Industry}}} 
      & P1 & 3.0 & 2.3 & 3.3 & 3.0 & 4.3 & 4.0 & \toolname \\
      & P2 & 4.3 & 2.0 & 5.0 & 1.6 & 4.3 & 5.0 & \toolname \\
      & P3 & 3.6 & 2.3 & 3.6 & 2.6 & 2.0 & 2.3 & \toolname \\
      & P4 & 2.6 & 1.3 & 3.3 & 1.0 & 4.3 & 1.0 & \toolname \\
      & median & \textbf{3.3} & 2.3 & \textbf{3.3} & 2.1 & \textbf{4.3} & 3.1 & \toolname \\
    \bottomrule
    \parbox[t]{2mm}{\multirow{5}{*}{\rotatebox[origin=c]{90}{Students}}} 
      & P5 & 2.0 & 2.6 & 3.3 & 3.0 & 4.0 & 3.6 & \toolname \\
      & P6 & 3.0 & 3.3 & 2.6 & 3.6 & 3.6 & 2.6 & \ddiff \\
      & P7 & 2.6 & 3.3 & 3.6 & 3.3 & 3.6 & 4.0 & \toolname \\
      & P8 & 2.0 & 3.6 & 3.0 & 4.3 & 2.0 & 3.0 & \ddiff \\
      & median & 2.3 & \textbf{3.3} & 3.1 & \textbf{3.4} & 3.6 & \textbf{3.8} & -- \\
    \bottomrule
\end{tabular}
\end{table*}

\revision{\autoref{tab:user_study_res} summarizes the results of the user study. As shown in the table, P1-P4 are the participants from industry and P5-P8 are the student participants. For each participant, the table shows the average score that the participant gives to the reports of each tool per each considered criterion. As explained in \autoref{sec:rq4_protocol}, three criteria are considered: ``Usefulness'', ``Clarity'', and ``Novelty''. For each criteria, the ``CS'' column represents the scores given to \toolname and ``DD'' represents the scores given to \ddiff. The last column on the right shows the  participant preference overall per the discussion at the end of the study. Note that the average score per participant is computed for the scores for which both \toolname and \ddiff give an output: C1-C3 for participants from industry and C5-C7 for students. This means the tools are tested on the same set of commits per participant group and a comparison between their scores is fair.}

\revision{For example, consider participant P1. This participant has given \toolname average scores of 3.0, 3.3, and 4.3 for usefulness, clarity, and novelty, respectively. All these scores are greater than or equal to 3, which means this participant's view about \toolname is more positive than negative with respect to all criteria. The average scores given to \ddiff by this participant are 2.3, 3, and 4.0 for usefulness, clarity, and novelty, respectively. This means this participant prefers \toolname in terms of all criteria, as they have given a higher average score to \toolname for each of them. Finally, P1 prefers \toolname overall as shown in the last column.}

\revision{All the other median scores for \toolname are above 3, except for median usefulness score given by students. This indicates the positive attitude of participants about the tool. Regarding the median usefulness score by students, we ask the participants what can be improved to make the augmentation more useful. Per our discussion, we notice that for two of the commits (C5 and C7) they also want to see the type of the variable that is taking a unique value in one version. This is a concrete suggestion for future improvement of our tool.}

\revision{We also note that median score that industry participants give to \toolname is at least one point higher that the median score they give to \ddiff in terms of each criteria. This indicates that people from industry find \toolname more helpful. Per our discussion with these participants (P1-P4), we realize that they think using \toolname is more practical, as the data it represents is more concise and its UI is integrated into GitHub. In contrast, \ddiff presents all values for all variables, which takes too long to understand. For the same reasons, all participants from industry prefer \toolname as mentioned in the last column of \autoref{tab:user_study_res}.}

\revision{Students give a higher score to \ddiff in term of usefulness. As mentioned above, the reason is that for two of the analyzed commits they want to see the type of the variable with unique value as well. However, in terms of clarity and novelty, they give the tools very close scores. The close score for the two tools is also reflected in the last column for students, as two people prefer \toolname (P5 and P7), and two prefer \ddiff (P6 and P8).}

\revision{In the live discussion after the scoring task, we ask the participants what they think about the idea of augmenting code diffs with runtime data. All participants have a very positive view on this idea. The positive view comes from the fact that these tools can ``ease the painful process of code review'' as P4 says, and ``help [us] catch unexpected behavior'' as P2 states. This all indicates there is a promising future for code diff augmentation with runtime data.}

\revision{We also ask the participants to score the augmented diff for commits that only \toolname can detect its runtime difference: C4 and C8. As a result, participants from industry give a median score of 3, 5, and 2 to C4 in terms of usefulness, clarity, and novelty, respectively. Students also give a median score of 4, 4.5, and 2 to C8 in terms of usefulness, clarity, and novelty, respectively. The novelty score is low for these two commits because these commits are adding/modifying literal strings, which means the runtime value for the invovled objects can be seen just by looking at the static code diff.}

\begin{mdframed}\noindent
    \revision{\textbf{Answer to RQ4: \rqfour} \\
    We have performed a user study involving four participants from industry and four students, who all scored \toolname and \ddiff outputs.
    Six of the eight participants prefer \toolname and seven participants have a positive view about code diff augmentation with runtime data.}
\end{mdframed}

\section{Discussion}
\label{sec:threats}

\subsection{External Validity}
\textit{Effectiveness on different targets:} In this paper, we focus on code changes that only modify one method of a Java program. Accordingly, as explained in \autoref{sec:rq1_protocol}, we evaluate the effectiveness on \dssize single-method code diffs from \ds. This is a threat to the external validity of this study, as it means the effectiveness results may not be generalizable to bigger changes in Java programs. Beyond size, we emphasize that we carefully build a diverse dataset, namely \rqthreeds, for the RQ2 and RQ3 experiments, from 10 various sizable Java projects. The satisfactory performance of \toolname on this dataset indicates the arguably wide applicability of \toolname on complex Java programs.

\subsection{Internal Validity}
\textit{Certainty of manual analysis results:} As described in \autoref{sec:rq3_protocol}, we conduct a manual analysis to assess the usefulness of \toolname augmented diffs. This manual analysis may lead to results that are subjective or uncertain. To address this, two participants with expertise in Java programming both perform the analysis and compare their assessments. In case of discrepancies, they have a thorough discussion about the results until they make a final conclusion. This protocol is meant to mitigate the threats created by the subjective nature of manual analysis.

\revision{\subsection{Code Changes with No Augmentation}}

\revision{There are two types of code diffs that are never augmented by \toolname. First, code diffs that do not cause any behavioral changes on matched lines. For example, refactoring the program by renaming a variable does not cause runtime differences. Second, there are code diffs that may cause some behavioral changes but do not produce any new state values. This happens when the code diff only changes the order in which state values appear, while \toolname works with set-based abstractions.}

\begin{lstlisting}[float=tb, style=diff, caption={\revision{A code diff that does not produce any new program state.}}, captionpos=b, label=lst:unimpactful_code_diff,belowskip=-0.4\baselineskip]
void print100() {
%\RHilight%  for (int i = 1; i <= 100; i++) {
%\GHilight%  for (int i = 100; i >= 1; i--) {
    System.out.println(i);
  }
}
\end{lstlisting}

\revision{For example, \autoref{lst:unimpactful_code_diff} shows the \texttt{print100()} method that runs a for loop and prints the natural numbers less than 100. The code diff in this example reverses the for loop iteration. In this original version (line 2), \texttt{i} goes from 1 to 100 and in the patched version (line 3), it goes from 100 to 1. Consequently, in the patched version, line 4 prints the numbers in reverse order. Even though the patch changes the behavior of the program as explained, the set of state values at the matched (line 4) is the same, they just appear in a different order. Therefore, \toolname does not catch the runtime difference in this code diff.}

\revision{The experiments in this study show that many of the important code diffs, such as bug-fixes, produce runtime differences going beyond value ordering, and are consequently augmented by \toolname.}

\revision{\subsection{Diff in Runtime Generated Code}
In Java, there are different options to generate and add code to the program at runtime, such as dependency injection, annotation processing and metaprogramming \cite{pawlak:hal-01169705}. In \toolname, the code added at runtime is not considered, this is because \toolname adds the breakpoints before the execution, by statically matching source code lines and comparing the two versions, as explained in \autoref{sec:breakpoint_identification}. Therefore, \toolname considers a runtime difference caused by the code added at runtime only if the difference is reflected in the state values in matched lines, later in the execution. As future engineering work, one can improve  \toolname by considering the code added at runtime, with sophisticated runtime analysis. This however poses UI challenges with respect to presenting the information to the developers in the diff.
}

\revision{\subsection{Scalability of \toolname}}
\revision{In the experiments of this paper, we use \toolname to augment code diffs that change a single method in different programs. In answer to RQ2, we see this already applies to a significant number of real-world commits. However, to make \toolname as applicable as possible and run it on all types of commits, incl. commits that change many different methods, future research will study scalability in terms of time and memory resources it needs. As shown in answer to RQ1 and RQ2, facing time limits is very rare for \toolname. This is because it just adds an extra computation statement at each line to extract state values, which does not change the order or magnitude of the execution time complexity. However, there are a notable number of code changes studied in RQ1 and RQ2 on which \toolname fails due to exceeding the memory limit. In this regard, we first note that the state depth configuration provided by \toolname helps users configure the tool to get the most detailed trace, while avoiding memory failure. Nevertheless, our careful inspection of the experiments reveal that we can avoid memory failures even with a very high state depth by making one technical improvement: \toolname could write and read the execution trace in small batches. This engineering improvement would enable \toolname to compute runtime differences without saving the whole trace in RAM.}

\revision{\subsection{Test Suite Inputs}}
\revision{As mentioned in \autoref{sec:appraoch_input}, \toolname needs a differentiating test as input to be able to detect runtime differences. This means \toolname does not augment commits with low quality test suites. For example, in our search for real-wolrd commits in the \rqthreeds dataset, we see that we collect 212 commits and 26\% (56/212) of them satisfy all the requirements except having a test that covers the code diff. Moreover, in \autoref{sec:rq3_results}, 18 commits have tests that covers the code diff but none of them shows a runtime behavioral difference. This entails that a significant number of real-world commits do not have a strong enough test suite that includes a differentiating test. A number of tools have been proposed to amplify test suites and generate such differentiating tests \cite{danglot2019automatic}. In the future, \toolname can take advantage of such tools to find a runtime difference for commits without a strong test suite.}

\section{Related Work}
\label{sec:related}
As explained in \autoref{sec:terminology}, previous studies consider different segments of the execution data as part of the program state, from all variable values \cite{zeller2002isolating,pham2017learning} or some of the variable values \cite{huang2007automated,gupta2020deep,schaf2013explaining,xie2004checking,abou2018substate,mehne2018accelerating} to the call stack and program counter \cite{xu2007efficient}.

In the following, we first review studies on collecting execution data and using this data for detecting the behavioral difference between two versions of a program. Next, we look at works focused on improving code diffs by augmenting them with execution data.

\subsection{Collecting Execution Data}
Lewis \cite{lewis2003debugging} defines the concept of omniscient debugger as a tool that ``works by collecting events at every state change and every method call in a program''. Different tools have been created that fully or partially implement this concept of omniscient debuggers \cite{matsumura2014repeatedly}. The state-of-the-art omniscient debugger is introduced by Shimari et al. \cite{shimari2021nod4j,shimari2019near}. They introduce \textsc{SELogger}, which records events, such as method entry, method exit, return values, and reading or writing a value into a variable. In contrast with \toolname, \textsc{SELogger} does not record values that are stored deep inside visible program objects. This means, compared to \textsc{SELogger}, the logging mechanism of \toolname provides more information for identifying execution differences.

Previous studies have also proposed execution trace collection tools for other programming languages, such as C/C++. Millnert et al. \cite{millnert2021dmfe} propose \textsc{DMCE} which adds probe statements after changed lines of C/C++ programs. After executing the code diff, they check which probe statements are executed. Based on the probe statement execution data, they determine which parts of the code diff are covered.

Magalhaes et al.\cite{magalhaesautomatic} introduce \textsc{Whiro}, which collects program state information in C/C++ programs. \textsc{Whiro} is built based on LLVM compiler and instruments the program at the compiler intermediate representation level. This tool can be customized to work at different levels of granularity. It can collect data allocated statically, in the stack or in the heap. Also, it can be set up to save the graph of relations between pointers up to a specified depth. Finally, it can be set up to collect data at different points: only before the main return, only before function returns, or after each data store statement. What makes Whiro novel and useful is how it collects data in an uncooperative environment, like in C/C++, where the program data does not have a datatype attached to it. It has to find the relation between low-level data at the heap and high-level program variables to do so. In contrast with \textsc{DMCE} and \textsc{Whiro}, \toolname collects execution data for Java programs.

\revision{Orton and Mycroft propose \textsc{Rehype} which collects the executed methods of a Java program \cite{orton2021refactoring}. Based on the collected data they output potential inefficient code in the program. They also introduce \textsc{Scopda} which uses \textsc{Rehype} output and suggests concrete improvements to the program \cite{orton2021source}. These two tools work with execution traces of Java programs, however their goal is different from \toolname: \toolname uses execution differences to explain code diffs, while \textsc{Scopda} analyzes execution trace to improve the efficiency of programs.}

\subsection{Execution Differencing}
Execution logs are one of the sources that can be used to model the execution behavior \cite{gupta2018runtime}. Goldstein et al. \cite{goldstein2017experience} present an algorithm to visualize behavioral differences using execution logs in four steps. First, they remove noisy lines of the logs that do not play an important role related to the behavior. Second, they create a finite state automaton (FSA) for each original/patched log using the \textsc{KTails} algorithm \cite{biermann1972synthesis}. Third, they find the differences between FSAs and finally, visualize the difference. This algorithm has been used in real-world setups in later studies \cite{deknop2021scalable,deknop2021log}. These studies show the effectiveness of Goldstein et al.'s algorithm in helping developers identify program changes. The main difference between \toolname and these works is that in contrast with \toolname, they do not consider state values to detect execution differences. This makes \toolname able to identify more fine-grained differences compared to these log-based differencing approaches.

Test suites may not execute programs exactly the same as they are executed in the field \cite{wang2017behavioral,javiertrace}. Nevertheless, it is common to utilize the traces produced during the execution of test suites to understand program changes \cite{wang2017behavioral,sherwood2008reducing}. Some researchers have introduced novel methods to generate tests that can differentiate between two versions of a program. Danglot et al. \cite{danglot2020approach} propose \textsc{DCI}. This approach amplifies existing tests to generate new tests that can detect behavioral changes in the continuous integration pipeline. Johnson et al. \cite{johnson2020causal} also use \textsc{EvoSuite} \cite{fraser2012whole} to generate tests for a given program. Then, they find two similar tests that one of them passes and the other one fails on the program. These tests are utilized to detect the root causes of defects in a program. These works are complementary to \toolname. They can create tests that can be used as the input covering test in \toolname.

Many studies compare execution traces to perform different tasks, such as fault localization \cite{guo2006accurately,wang2019explaining} and failure root cause analysis \cite{zhang2019inflection,yi2015synergistic}. The most advanced techniques for comparing execution traces are \textsc{TraceSim} \cite{rodrigues2022tracesim} and \textsc{S3M} \cite{khvorov2021s3m}. These two works compare stack traces that are collected as ordered lists of lines that are executed before a program crash. \textsc{TraceSim} performs the comparison by computing measures such as edit distance and tf-idf between traces and their elements. On the other hand, \textsc{S3M} uses deep learning to determine the similarity between two traces. For this, they present traces as vectors and then utilize the siamese architecture \cite{chopra2005learning} to compute the similarity between them. All these works take the list of executed statements as the execution trace, while \toolname also considers state values. This can make \toolname more precise in detecting differences in executions.

There are also several studies that compare variable values to detect what causes differences between two executions \cite{zeller2002isolating,zeller2002simplifying,cleve2005locating}. Cleve and Zeller \cite{cleve2005locating} use \emph{delta debugging} \cite{zeller1999yesterday,zeller2002simplifying} to narrow down differences between program states in two executions of a program. The result is a small set of variable value differences that make one execution of a program pass and another execution fail. Other researchers improve the precision of delta debugging by mixing it with other techniques \cite{yi2015synergistic}, such as execution coverage \cite{yu2012practical}, observation-based slicing \cite{binkley2013observation}, dual slicing \cite{sumner2013comparative}, and hierarchical information \cite{misherghi2006hdd}. \toolname is different from these tools mainly because it is focused on improving the code diff by adding a concise execution difference between two versions of the program. On the contrary, these tools are designed to identify the root cause of a failure.

\revision{Abramson et al. have introduced \emph{relative debugging} \cite{sosivc1997guard,abramson1996debugging,abramson2010data,dinh2013scalable,abramson2009relative,searle2003duct}. Relative debugging is implemented in their tool, called \textsc{Guard}. \textsc{Guard} runs two programs and reports the differences between the values assigned to their variables at runtime. This is close to what \toolname does, however \textsc{Guard} requires the users to specify the data structures, the variables, the breakpoints, and the mappings between the variables of two versions. In \toolname, these steps are all done automatically. Consequently, \toolname is able to generate one augmentation for a given code diff, while \textsc{Guard} acts similar to a traditional debugger, in which the user interacts with the tool to collect runtime differences and explore them.}

The tool that is most similar to \toolname is \textsc{didiffff} \cite{kanda2022didiffff}. As explained in \autoref{sec:baseline}, \textsc{didiffff} lists values assigned to each variable during execution and compares the lists for two versions of a program. \textsc{didiffff} is built on top of \textsc{SELogger}. There are two main differences between \toolname and \textsc{didiffff}. \toolname explores the data inside complex Java objects, while \textsc{didiffff} only considers values of primitive variables. Also, \toolname only reports a concise execution difference in its UI output, while \textsc{didiffff} reports all the extracted differences.

\begin{table*}[t]
\centering
\scriptsize
\caption{\revision{\toolname compared to the related work.}}
\label{tab:related_works}
\begin{tabular}{l l c c c c}
\toprule
	Tool & Language & Collected Runtime Data & Computed Runtime Diff & Presentation & Open-source \\
	\midrule
	\textsc{SELogger}~\cite{shimari2021nod4j} & Java & Variable \& Return Values & -- & Log File & \cmark \\
    \textsc{DMCE}~\cite{millnert2021dmfe} & C/C++ & Covered Lines & -- & Log File & \cmark \\
    \textsc{Whiro}~\cite{magalhaesautomatic} & C/C++ & Heap Data & -- & Graphical Visualization & \cmark \\
    \textsc{VISSOFT}~\cite{deknop2021scalable} & -- & -- & Detects Outlier Logs & Graphical Visualization & \cmark \\
    \textsc{DCI}~\cite{danglot2020approach} & Java & Covered Lines & Coverage Difference & Amplified Test & \cmark \\
    \textsc{TraceSim}~\cite{rodrigues2022tracesim} & -- & Stack Trace & Stack Trace Similarity & Similarity Score & \cmark \\
    \textsc{DeltaDebugging}~\cite{zeller1999yesterday} & C/C++ & Variable Values & Failure-inducing Changes & Text & \cmark \\
    \textsc{Guard}~\cite{sosivc1997guard} & C/Fortran & Variable Values & Variable Value Difference & Text \& Visualization & \xmark \\
    \ddiff~\cite{kanda2022didiffff} & Java & Variable Values & Variable Value Difference & Ad hoc GUI & \cmark \\
    \toolname (this paper) & Java & State \& Return Values & State \& Return Value Diff & Augmented Code Review Diff & \cmark \\
    \bottomrule
\end{tabular}
\end{table*}

\subsection{Improving Code Diffs}
Previous studies have shown how integrating runtime data into the developers' working environment helps them better perform tasks, such as bug fixing and code comprehension \cite{rothlisberger2010augmenting,liu2019augmenting}. In this section, we review studies focused on improving code diffs and making them more understandable. This can be done in different ways, such as splitting code changes into smaller sets of related changes \cite{dias2015supporting} or by improving how code changes are presented. Our focus is on the diff presentation.

Some tools improve the code diff presentation by making it more fine-grained, such as \textsc{Mergely} \cite{mergely} and \textsc{GumTree} \cite{falleri2014fine}. For this, \textsc{Mergely} considers changes at the level of code elements inside lines, instead of looking at whole line differences. \textsc{GumTree} enhances existing AST differencing algorithms, such as \textsc{ChangeDistiller} \cite{fluri2007change}, and proposes an algorithm that computes minimum changes that should be made on the original AST to reach the patched one. \textsc{GumTree} marks these fine-grained AST level changes on the code diff to help developers easily see what is the exact change to the program.

Decker et al. \cite{decker2020srcdiff} introduce a new syntactic differencing tool, called \textsc{srcDiff}. This tool is built on top of \textsc{srcML} which represents source code in an XML format and annotates the code with the syntax information. Based on manual and statistical analysis, they propose several heuristics to determine if a change is actually a modification or a complete replacement. Hence, in contrast with \textsc{GumTree}, the goal of \textsc{srcDiff} is not producing an optimal diff. \textsc{srcDiff} intends to produce a diff that is more similar to the changes performed by developers. The difference between \toolname and tools like \textsc{GumTree} and \textsc{srcDiff} is that \toolname adds execution data to the diff, while they solely use static data to improve the code change presentation.

\revision{Another way of improving code diff presentation is based on the idea of literate programming. \texttt{literate-diff-viewer}\cite{Abagames} is a tool where the description of the source code changes is provided in natural language and then the diff generated shows how the application behavior varies, together with the description. Applied to video game diffs, one see how the games evolve as we scroll through the diff description.}

In another line of code, researchers try to integrate new information into the integrated development environments (IDE) \cite{sulir2018visual}. This can include putting code change information with data related to developers' actions in the IDE \cite{proksch2017enriching} or production data \cite{winter2019monitoring}. For example, Cito et al. \cite{cito2019interactive} add the time spent to run a method at production to the IDE code editor. When the code is changed, they predict the new production time for the changed code and show it to the developer in IDE. As a result, developers better detect performance problems in their code changes. These works are different from \toolname as they do not add program state data to the code diff.

Bohnet et al. \cite{bohnet2009projecting} combine execution traces with code changes to help developers identify the root cause of a failure in C/C++ programs. They define the execution trace as the sequence of executed functions. When a program faces a failure after a code change, the execution trace is collected and shown to the developer. Next, the developer interacts with the tool to detect the failure causing behavioral change. This work is different from \toolname as it does not consider state values as a part of the execution trace. In contrast with \toolname, it is also a semi-automatic tool requiring interaction with the developer.

\revision{\autoref{tab:related_works} summarizes the closest related work to \toolname. \toolname is the first open-source execution differencing tool for Java that explores deep inside the value of all non-primitive types and computes state values.} \toolname detects and reports a unique state value in the original or patched version of a program to help developers understand the changed behavior.

\section{Conclusion}
\label{sec:conclusion}
In this paper, we introduce \toolname, a novel tool that detects runtime differences between two versions of a program and augments the code diff with it. The augmentation is done with unique variable and return values that occur during the execution of a test case. This new information is important for developers to better understand behavioral changes caused by a code change. We demonstrate the effectiveness of \toolname by running it on \dssize code diffs for Defects4J bugs. \revision{\toolname produces an execution difference for 95\% (555/\dssize) of the code diffs, better than the closest related work. The high quality of \toolname's augmented diffs is also confirmed by a manual analysis of 30 commits from real-world open-source projects. Our user study shows the participants find the idea of code diff augmentation with runtime useful for software development process.}

In the future, research is needed to detect and mitigate spurious program state differences. Such research relates to the currently active research field on flaky tests. Moreover, while this work focuses on bug fixes, the augmentation done by \toolname could also help explain other types of changes, e.g., refactorings. Additional studies are required to investigate this potential application. Finally, \toolname could be integrated into a fully-automated pipeline, e.g., continuous integration, so that developers receive its augmented code diff directly after proposing changes for code review.

\revision{
\section{Acknowledgments}
\label{sec:acknowledgments}
We acknowledge the important contribution of Julian Wachter to the engineering of \toolname in the late stage of the project.
This work was partially supported by the Wallenberg Artificial Intelligence, Autonomous Systems and Software Program (WASP) funded by Knut and Alice Wallenberg Foundation, and by the Swedish Foundation for Strategic Research (SSF). Some experiments were performed on resources provided by the Swedish National Infrastructure for Computing (SNIC).
}

\bibliographystyle{IEEEtran}
\bibliography{references}

\end{document}